\pdfoutput=1
\documentclass[10pt,a4paper,3p,final]{elsarticle}
\usepackage[latin9]{inputenc}
\usepackage{textcomp}
\usepackage{amsmath}
\usepackage{amssymb}
\usepackage{graphicx}
\usepackage{esint}
\usepackage{subscript}
\usepackage[unicode=true,
 bookmarks=true,bookmarksnumbered=false,bookmarksopen=false,
 breaklinks=false,pdfborder={0 0 1},backref=false,colorlinks=false]
 {hyperref}
\hypersetup{pdftitle={Spectral properties of transition metal pnictides: theory and experiment},
 pdfauthor={A. van Roekeghem et al}}

\makeatletter

\pdfpageheight\paperheight
\pdfpagewidth\paperwidth

\makeatother

\begin{document}

\title{\textbf{\Large Spectral properties of transition metal pnictides
and chalcogenides: angle-resolved photoemission spectroscopy and dynamical
mean field theory}}

\author{Ambroise van Roekeghem}

\address{Centre de Physique Théorique, Ecole Polytechnique, CNRS, Université
Paris-Saclay, 91128 Palaiseau, France}

\address{Beijing National Laboratory for Condensed Matter Physics, and Institute
of Physics, Chinese Academy of Sciences, Beijing 100190, China}

\ead{vanroeke@cpht.polytechnique.fr}

\author{Pierre Richard, Hong Ding}

\address{Beijing National Laboratory for Condensed Matter Physics, and Institute
of Physics, Chinese Academy of Sciences, Beijing 100190, China}

\address{Collaborative Innovation Center of Quantum Matter, Beijing, China}

\author{Silke Biermann}

\address{Centre de Physique Théorique, Ecole Polytechnique, CNRS, Université
Paris-Saclay, 91128 Palaiseau, France}

\address{Collège de France, 11 place Marcelin Berthelot, 75005 Paris, France}

\address{European Theoretical Synchrotron Facility, Europe}
\begin{abstract}
Electronic Coulomb correlations lead to characteristic signatures
in the spectroscopy of transition metal pnictides and chalcogenides:
quasi-particle renormalizations, lifetime effects or incoherent badly
metallic behavior above relatively low coherence temperatures are
measures of many-body effects due to local Hubbard and Hund's couplings.
We review and compare the results of angle-resolved photoemission
spectroscopy experiments (ARPES) and of combined density functional
dynamical mean field theory (DFT+DMFT) calculations. We emphasize
the doping-dependence of the quasi-particle mass renormalization and
coherence properties.\end{abstract}
\maketitle

\section{Introduction}

Angle-resolved photoemission spectroscopy (ARPES) is a powerful tool
to probe the electronic properties of materials. It allows for a measurement
of the momentum-resolved spectral function, providing crucial information
on the Fermi surface, the quasi-particle dispersions, or even the
momentum-resolved magnitude of the superconducting gap. It can also
give insights into the orbital characters of the electronic states
through the use of polarized light, core and plasmon excitations,
and some clues about the lifetime of quasiparticles -- though the
latter point is more delicate since the measurements can be influenced
by the sample quality or other extrinsic limitations. In the following,
we will focus on measurements in the paramagnetic normal state, except
for Section \ref{sec:SC gap} where we summarize recent work on measurements
of the superconducting gap.

This review is organized as follows: in Sections 2 and 3 we give brief
introductions to the basics of photoemission spectroscopy and dynamical
mean field based electronic structure calculations respectively. Section
4 provides an overview over the spectral properties of transition
metal pnictide and chalcogenides from a combined experimental and
theoretical point of view. Section 5 is devoted to measurements of
the superconducting gap, while Section 6 summarizes recent theoretical
advances.

\section{Angle-resolved photoemission spectroscopy}

\label{sec:ARPES}

\subsection{Introduction}

Photoemission spectroscopy, also called photoelectron spectroscopy,
is a technique based on the photoelectric effect \citep{Hertz-1887}.
A light source is used to produce photons of a given energy, which
are sent on a sample with a chosen incidence. An electron of the sample
can then absorb a photon of the incident beam and escape with a maximum
energy $E=\hbar\omega-\phi$ with $\phi$ the workfunction of the
material. By collecting those electrons and analyzing their energies
and wave vectors, one obtains information about the electronic structure
of the material, such as a direct visualization of the quasiparticle
dispersions and of the Fermi surface (for a comprehensive book, see
e.g. Ref. \citep{Hufner}, for an introduction, see e.g. Ref. \citep{Damascelli-ARPES,Richard_JPCMreview}).

\begin{figure}[h]
\begin{centering}
\includegraphics[width=8cm]{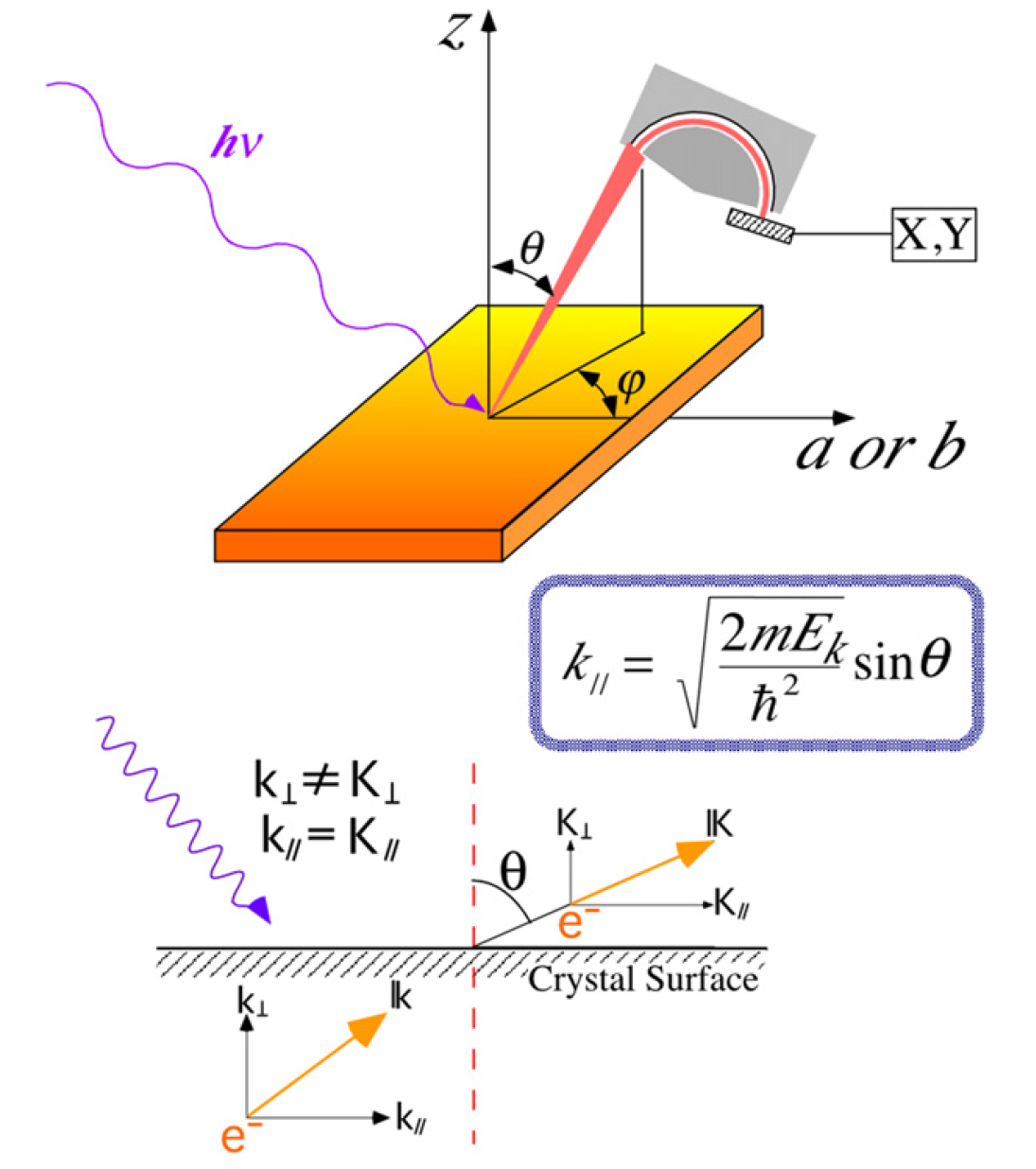}
\par\end{centering}

\caption[Principles of an ARPES experiment]{Principles of an angle-resolved photoemission experiment (reprinted
with permission from Ref. \citep{Pierre-ARPES-review}, copyright
\copyright (2011) by IOP Publishing).\label{fig:ARPES}}
\end{figure}

The photoemission process is often interpreted within the three-step
model, in which the electron is excited from an initial state to a
final state, then travels to the surface, and finally escapes from
the solid. Fundamentally, the analysis of the photoemission spectrum
is a many-body problem, for the escape of an electron leaves the solid
in an excited state that may involve several electrons. Another description
making use of fewer approximations is the one-step model, in which
the photoemission process is described as an optical transition between
a ground-state many-body wave function involving N electrons and an
excited wave function involving N-1 electrons and an escaping plane
wave. The photocurrent is described within the sudden approximation,
which states that the creation of the photohole is instantaneous and
that there is no interaction between the escaped electron and the
remaining system. The photocurrent in the three-step model is proportional
to the probability of transition between the ground state and all
final states. For a given final state it is determined by Fermi's
golden rule: 
\begin{equation}
w_{fi}=\frac{2\pi}{\hbar}|<\psi_{f}^{N}|H^{int}|\psi_{i}^{N}>|^{2}\delta(E_{f}^{N}-E_{i}^{N}-h\nu)
\end{equation}

In the sudden approximation, $<\psi_{f}^{N}|H^{int}|\psi_{i}^{N}>=<\phi_{f}^{k}|\tilde{H}^{int}|\phi_{i}^{k}><\psi_{f}^{N-1}|c_{k}|\psi_{i}^{N}>$,
where $\phi^{k}$ is a one-electron wave function.

This factorization is valid in the limit of infinite photon energy,
and is used as an approximation in practice. It has been argued that
the adiabatic to sudden transition takes place at a very large photon
energy -- of the order of the keV -- when many-body effects like core
electron - plasmon interaction take place, while smaller energies
are required in the case of a localized system (see for instance \citep{Lee-adiabatic-sudden,Hedin-sudden}).
Still, most of the ARPES experimental analysis is nowadays done within
the sudden approximation. For a pedagogical discussion of effects
beyond the sudden approximation we refer the reader to Ref.~\citep{Berthod}.
A DMFT-based approach in the framework of the one-step model has been
worked out in Ref.~\citep{Minar_ARPES}.

Summing over all possible excited final states $\psi_{e}^{N-1}$ with
energy $E_{e}^{N-1}$ and writing $<\phi_{f}^{k}|\tilde{H}^{int}|\phi_{i}^{k}>=M_{f,i}^{k}$
the one-electron dipole matrix element, we see that the probability
of detecting an electron with wave vector $k$ and energy $E_{kin}$
is proportional to:

\begin{equation}
I(k,E_{kin})\propto\sum_{f,i}|M_{f,i}^{k}|^{2}\sum_{e}|<\psi_{e}^{N-1}|c_{k}|\psi_{i}^{N}>|^{2}\delta(E_{kin}+E_{e}^{N-1}-E_{i}^{N}-h\nu)
\end{equation}

Finally, we take into account the resolution of the experiment and
the effect of temperature and see that the photocurrent is proportional
to the spectral function $A(k,\omega)=\sum_{e}|<\psi_{e}|c_{k}|\psi_{0}>|^{2}\delta(\omega+E_{e}^{(N-1)}-E_{0}^{(N)}),\omega<0$:

\begin{equation}
I(k,\omega)=I_{0}(k,\nu,\overrightarrow{A})A(k,\omega)f(\omega)\ast R(k,\omega)\label{eq:Photocurrent}
\end{equation}
 with $R$ being the resolution function of our ARPES experiment,
$f$ the Fermi function, and $I_{0}$ a prefactor depending on one-electron
matrix elements. $\omega$ is the energy of electrons in the solid
with respect to the Fermi level. The spectral function is a quantity
we calculate directly from first principles using Dynamical Mean Field
Theory (DMFT), see Section \ref{sec:First principles} and Ref. \citep{Tomczak-CRAS}
for an overview of calculations of spectral properties for oxide materials.
Hence, those two techniques are an ideal match to compare experimental
and theoretical results. However, due to the small mean free path
of electrons at the kinetic energies we consider \citep{Seah-Mean-free-path},
ARPES is a surface probe. This can further complicate the interpretation
of the results if the surface electronic structure is different from
the bulk. Furthermore, the component of the wave vector perpendicular
to the plane is not conserved during the photoemission process, only
the in-plane component is (see Figure \ref{fig:ARPES}).

Under the assumptions of an independent particle picture, the sudden
approximation and the three-step model, one can directly study the
momentum-resolved band-structure of the solid. Under these conditions,
one can also extract the perpendicular component of the wave vector:
using a nearly-free electron model for the final state, one has the
dispersion:

\begin{equation}
k_{\perp}=\frac{\sqrt{2m}}{\hbar}\sqrt{E_{kin}cos^{2}\theta+V_{0}}
\end{equation}
 with $\theta$ the polar angle as in Figure \ref{fig:ARPES} and
$V_{0}$ the inner potential, which depends on the material considered
and represents the bottom of a free-electron-like band dispersion
for the final state with respect to the vacuum level. In practice,
to estimate $V_{0}$ one can vary the photon energy -- which is possible
using synchrotron light -- and observe a periodic dispersion of the
band structure.

In recent years, important progress has been made both on the detection
setup and on the light sources. At present, ARPES can provide a resolution
of the order of the meV and access a large temperature range, even
below 1K. Using a laser as the light source allows for a very fine
resolution and a higher sensitivity to the bulk but limits the momentum
and energy range that can be accessed due to the small difference
between the photon energy and the workfunction. In contrast, synchrotron
light offers the possibility to tune the photon energy and thus to
select the wave number perpendicular to the sample surface, or to
use soft x-rays to also probe the bulk of the crystal and a large
area of the momentum space. The versatility of this technique is also
exceptional, since it can be used on materials synthesized in situ
or on a microscopically defined area, and with spin- or time-resolution.

An obvious limitation is of course that the information provided by
ARPES is limited to the occupied part of the spectrum. Inverse photoemission
techniques such as Bremsstrahlung isochromate spectroscopy (BIS) on
the other hand suffer from generally much lower resolution, due to
the weak cross section of the electron addition process \citep{Hufner}.
Pump-probe techniques where the electronic system is excited by a
pump pulse before being probed by ARPES provide a promising alternative.
Assuming that the main effect of the pump pulse is to heat the electronic
system to a high effective temperature, one obtains information on
empty states within the corresponding energy range \citep{Perfetti-insulator-metal,Perfetti-Mott-insulator}.
We note, however, that the conditions of validity of such an assumption
are a largely open question, since in general one faces a truly non-equilibrium
situation invalidating the notion of effective electronic temperature
and making the interpretation of the experiment much more involved.
In practical applications to iron pnictide compounds \citep{Avigo-electron-phonon,Rettig-electron-phonon,Torchinsky-pump-probe,Kumar-relaxation}
the main interest in applying pump-probe spectroscopy has rather been
lying in the possibility of analyzing different relaxation processes,
giving information e.g. on electron-phonon coupling.

\subsection{Conventions and notations}

\begin{figure}
\begin{centering}
\includegraphics[width=12cm]{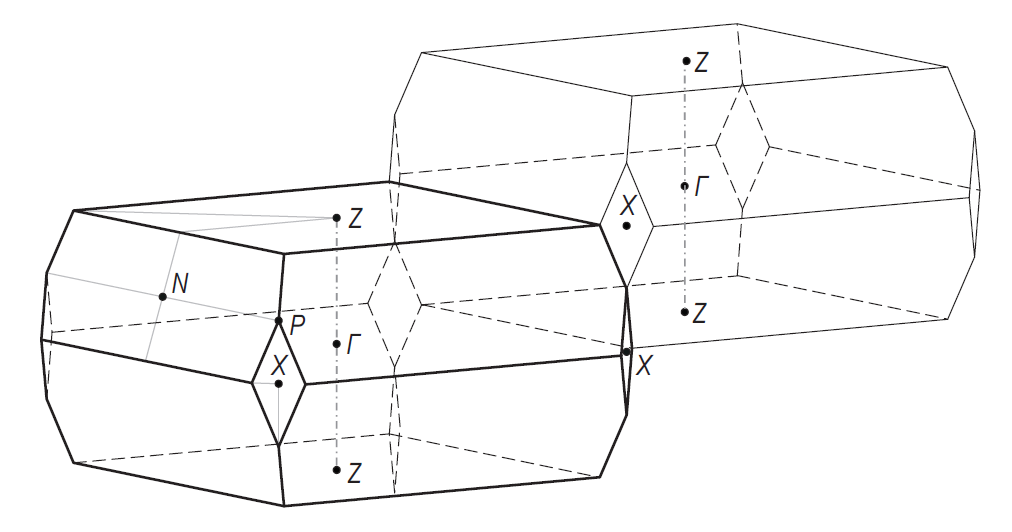} 
\par\end{centering}

\caption[Body-centered tetragonal Brillouin zone]{Body-centered tetragonal first and second Brillouin zone (reprinted
with permission from Ref. \citep{Andersen-Boeri}).\label{fig:Crystallo BZ}}
\end{figure}

Iron pnictides in the tetragonal, paramagnetic phase present two different
crystal structures: while the 11, 111 and 1111 have simple tetragonal
stacking with \textit{P4/nmm} space group, the 122 family crystallizes
in the body-centered tetragonal ThCr\textsubscript{2}Si\textsubscript{2}-type
crystal structure (\textit{I4/mmm} space group), in which the two
arsenic atoms of different layers are on top of each other -- the
layers are translated before they are stacked on top of each other.
The three-dimensional Brillouin zone of this family is represented
in Figure \ref{fig:Crystallo BZ}, with the crystallographic notations
for the high-symmetry points. The literature on iron pnictides can
be confusing because high-symmetry points are often denominated differently,
due to the different space groups and authors considering so-called
``one-Fe'' or ``two-Fe'' unit cells -- that is to say, a Brillouin
zone that ignores the translation-breaking potential created by the
alternating positions of the arsenic atoms above and below the iron
plane, or one which takes it into account. In particular, the crystallographic
X point in the 122 Brillouin zone is often named M in the ``one-Fe''
notation. In contrast, the corresponding point of the electronic structure
of the\textit{ P4/nmm} Brillouin zone is also named M in the crystallographic
conventions.

\begin{figure}
\begin{centering}
\includegraphics[width=8cm]{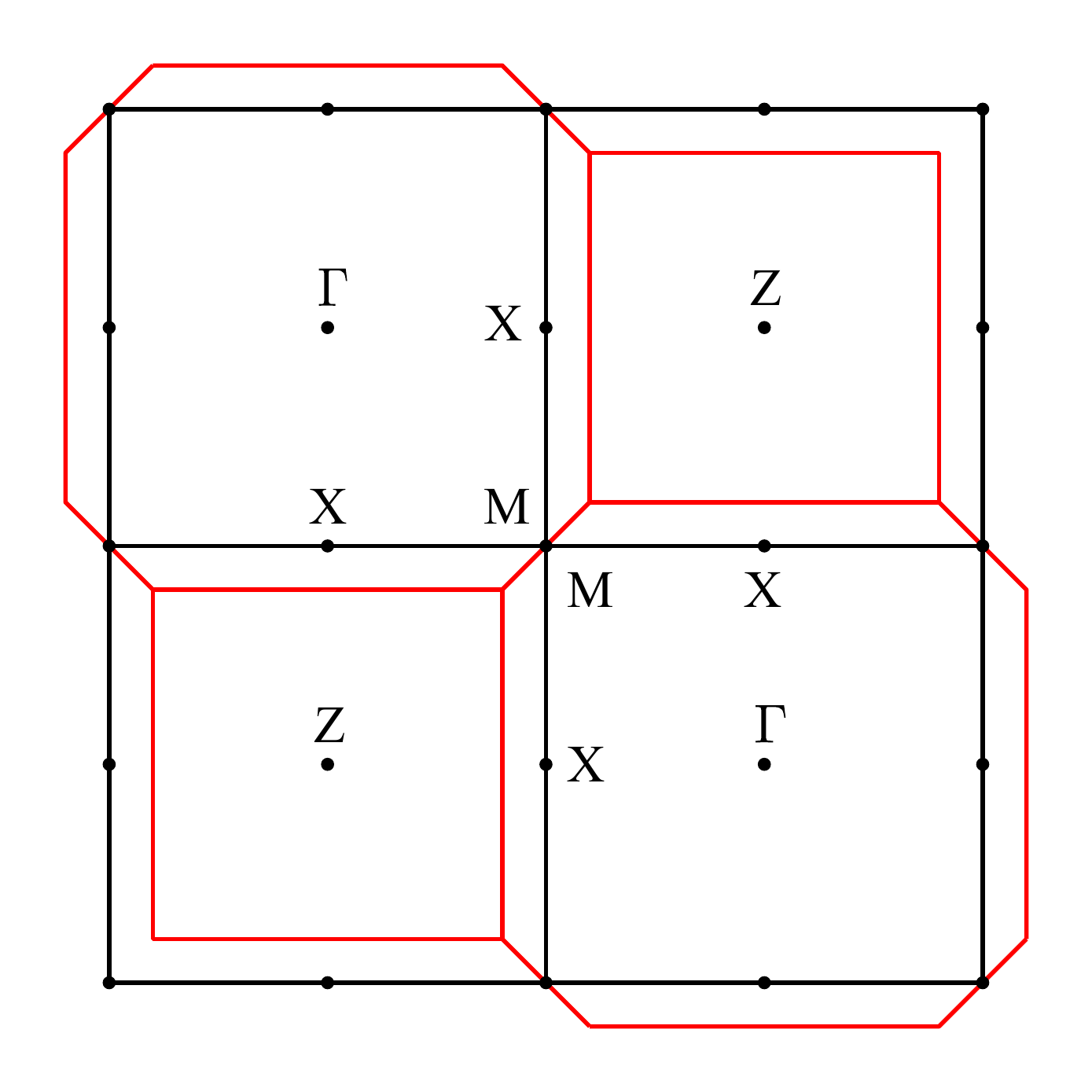} 
\par\end{centering}

\caption[Notations for the high-symmetry points]{Cut of the body-centered tetragonal Brillouin zones in the $k_{z}=0$
plane (red) along with our choice of notations for the high-symmetry
points (black). The simple tetragonal Brillouin zones would correspond
to black lines, but the Z points would be replaced by $\Gamma$ points.\label{fig:BZ notations}}
\end{figure}

In our case and for practical purposes, we will define our points
for the body-centered tetragonal Brillouin zone as in Figure \ref{fig:BZ notations},
which shows a cut of the Brillouin zone at $k_{z}=0$. One can see
that our X point is actually not a high symmetry point along the $\Gamma$XZ
direction -- where there is no high symmetry point a priori. On the
other hand, M is a well defined high symmetry point on both $\Gamma$M$\Gamma$
and ZMZ directions. Interestingly, in the simple tetragonal Brillouin
zone X becomes a high symmetry point along the $\Gamma$X direction,
for the Z points of Figure \ref{fig:BZ notations} would be replaced
by $\Gamma$ points. This choice allows for a relative uniformity
of notations within the different pnictides families, and is in line
with the more often encountered ``one-Fe unit cell'' notations.

\subsection{Orbital characters and matrix elements}

A common procedure in ARPES to obtain information about the orbital
characters of electronic states consists in using different light
polarizations to highlight or filter certain symmetries. The matrix
element $I_{0}(k,\nu,\overrightarrow{A})$ is proportional to: 
\begin{equation}
I_{0}(k,\nu,\overrightarrow{A})\propto|<\phi_{k}^{f}|\overrightarrow{\epsilon}\cdot\overrightarrow{x}|\phi_{k}^{i}>|^{2}
\end{equation}
 with $|\phi_{k}^{f}>$ and $|\phi_{k}^{i}>$ the final and initial
states respectively and $\overrightarrow{\epsilon}$ a unit vector
along the direction of the polarization of the vector potential $\overrightarrow{A}$.
The principle is then to find a mirror plane in which we detect the
photoelectrons that allows us to classify states into even or odd
symmetry with respect to this plane. In that geometry, the final state
has to be even to be detected, and so does $\overrightarrow{\epsilon}\cdot\overrightarrow{x}|\phi_{k}^{i}>$
in order for the matrix element to be non-zero. Finally, if $\overrightarrow{\epsilon}$
is in the mirror plane (p polarization) we obtain that $|\phi_{k}^{i}>$
has to be even, and if $\overrightarrow{\epsilon}$ is perpendicular
to the plane (s polarization) $|\phi_{k}^{i}>$ has to be odd.

\begin{figure}
\begin{centering}
\includegraphics[width=12cm]{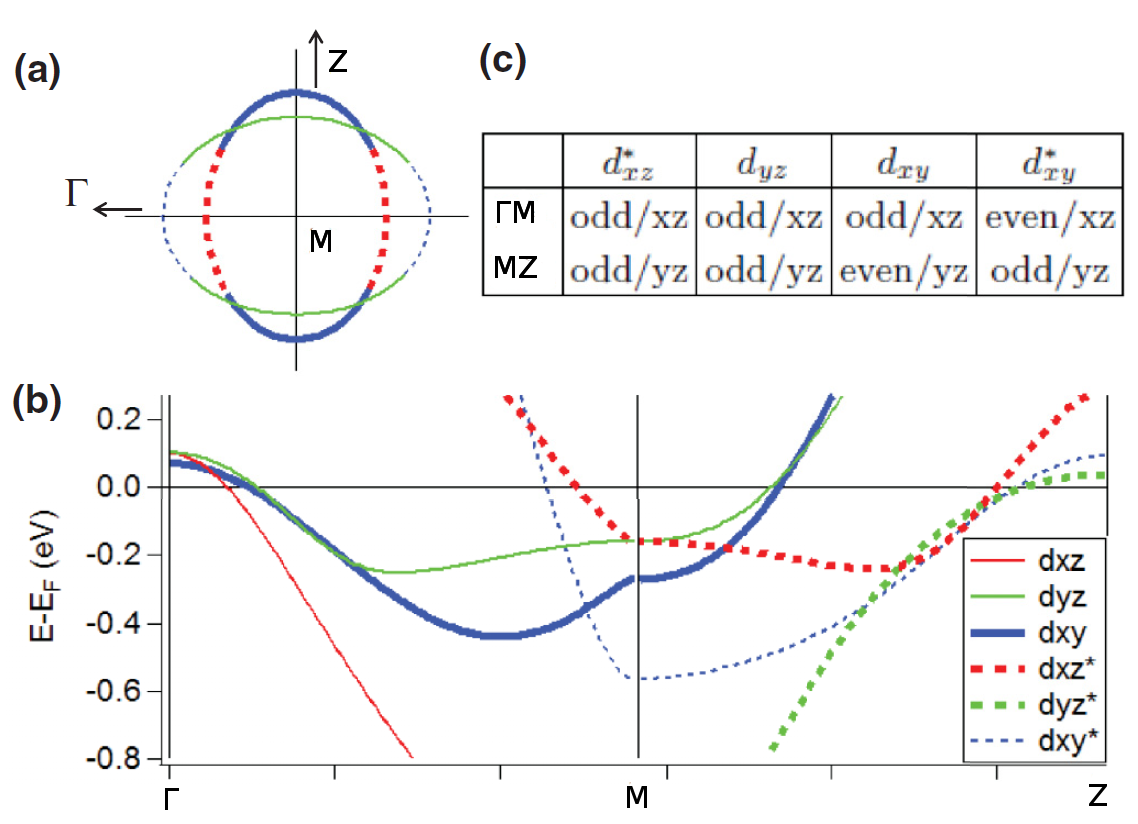} 
\par\end{centering}

\caption[Effect of the 2-Fe unit cell on matrix elements]{Effect of the 2-Fe unit cell on matrix elements\protect \protect
\protect \protect \\
 (a) Sketch of the bands forming the electron pockets. Solid and dotted
lines represent the in-phase and out-of-phase orbitals. Colors indicate
the main orbital character. (b) Sketch of the dispersion of the different
bands along the path $\Gamma$MZ (in the $k_{z}=0$ plane). The color
sketches the main orbital character, although there may be hybridization
with other orbitals. (c) Symmetries of the different bands forming
the electron pockets at M. (Adapted with permission from Ref. \citep{Brouet-2Fe}
(copyright \copyright (2012) by The American Physical Society), using
the notations of Figure \ref{fig:BZ notations})\label{fig:2-Fe unit cell}}
\end{figure}

However, in iron pnictides the problem is more complicated. Due to
the fact that the arsenic position is alternating above and below
the iron plane, creating a 2-Fe unit cell, there is no such mirror
plane containing the $\Gamma$M direction. Still, a possibility to
attribute symmetries to the electronic states is to consider this
alternating position of As as a small perturbing potential that breaks
the translational symmetry and to analyze the band structure as a
superposition of a ``main'' in-phase band structure in the sense
of the 1-Fe unit cell and of the corresponding out-of-phase folded
band structure \citep{WeiKu-2Fe-BZ,Brouet-2Fe}. The symmetry of the
electronic states is found to be different due to the alternating
arsenic position. Furthermore, a second effect appears, due to the
fact that the final state -- close to a free-electron state -- is
not sensitive to the potential breaking the translational symmetry.
As a consequence, the spectral weight is different in the first and
second Brillouin zones, with the bands originating mainly from folding
having little spectral weight in the first Brillouin zone. Those results
are summarized in Figure \ref{fig:2-Fe unit cell}, in particular
for the electron pockets around the M point, which are the most impacted.

\section{First-principles approaches to spectral properties of transition
metal pnictide materials}

\label{sec:First principles}

When studying the electronic properties of materials, theories assuming
a single-particle band picture where the electrons are considered
to be non-interacting fermions in an effective periodic potential
often produce remarkable results. In fact, for ground state properties
the Hohenberg-Kohn theorem of Density Functional Theory guarantees
the exact solution to be of this form, and Density Functional Theory
-- even though in practice using approximations to the exact functional
-- has been able to produce \textit{ab initio} quantitative predictions
for many chemical and physical systems \citep{Kohn-nobel}. However,
care is in order when dealing with excited state properties where
the Kohn-Sham band structure of DFT is not \textit{a priori} a reasonable
approximation. This is intrinsically true for the photoemission spectrum
where effects beyond the band picture are expected and found. In order
to arrive at a theoretical description that incorporates those into
the ab initio description one borrows concepts from many-body theory,
originally developed for lattice models of correlated fermions.

The key phenomenon modeled in interacting fermionic lattice models
is the competition between the tendency to delocalization of the electrons
driven by the corresponding lowering of the kinetic energy and local
Coulomb interactions counteracting this delocalization. The most simple
model -- but containing very rich physics -- was introduced by J.
Hubbard \citep{Hubbard_model_1963,Hubbard_model_1964}. The idea is
to include correlations in the original Hamiltonian by adding a Coulomb
repulsion when several electrons are on the same atomic site. While
the model was initially introduced in an \textit{ad hoc} manner, nowadays,
this interaction $U$ between two electrons can actually be evaluated,
for instance using the constrained-Random Phase Approximation (cRPA)
\citep{cRPA-ferdi-2004}: with $\chi$ the wave function associated
to the considered atomic level on the site $R$,

\[
U\thicksim\int drdr'|\chi_{R}(r)|^{2}W^{r}(r-r')|\chi_{R}(r')|^{2}
\]
 where $W^{r}$ is a partially screened interaction between electrons.
For more details on the calculation of $U$ and of its frequency dependence
in iron pnictides, see e.g. Ref. \citep{loig-tesis,vaugier-crpa,
Hubbard-interactions, Miyake2010}.

In the single-orbital case at zero temperature the Hubbard $U$ and
the hopping terms are the only energy scales of the system. The physics
is then simply determined by the competition between two phenomena:
the electronic correlations introduced by $U$ which tend to localize
the electrons and the kinetic delocalization characterized by the
hopping amplitudes. By tuning the ratio $U/W$, with $W$ the bandwidth,
one can then go all the way from the weak coupling (band) limit to
the strong coupling (atomic) limit. Among compounds with partially
filled 3\textit{d} or 4\textit{f} shells there are numerous examples
of materials in the intermediate regime, where spectra are not well-described
by the single-particle density of states nor solely consist of Hubbard
bands but exhibit a renormalized quasi-particle peak accompanied by
incoherent higher-energy excitations.

As discussed in Section \ref{sec:Spectral properties}, iron pnictides
are in the intermediate regime where it is necessary to use a theory
able to probe both strongly and weakly correlated regimes, such as
Dynamical Mean Field Theory (DMFT), even though the multiorbital nature
of the systems brings in additional interesting physics: the five
3\textit{d} orbitals of the transition metal have a total bandwidth
of about 5 eV within DFT, to be compared to a $U$ of about 3 eV.
In this -- \textit{a priori} moderately correlated -- regime, the
actual correlation strength is tuned by the suppression of screening
channels due to Hund's rule: the Coulomb cost for two electrons with
anti-parallel spins is larger than the one for two electrons with
parallel spins, since in the latter case the Pauli principle keeps
the electrons apart. This intra-atomic exchange, the Hund's rule coupling
$J$, leads to most interesting multiorbital physics in this class
of compounds \citep{HauleNJP11,udyn-werner,Medici-selective-Mott}.

Over the last decades, DMFT has been at the origin of impressive success
in providing us with a qualitative understanding but also quantitative
spectral functions directly comparable to experiments. We present
below a summary of the basic principles of DMFT (for extensive reviews,
see \citep{biermann_ldadmft,LDA-DMFT-biermann,RevDMFT_AG,georges-DMFT-2004,kotliar-review-DMFT,bulla-DMFT})
and of its combination with DFT for the case of transition-metal pnictides.

\subsection{Combination of DFT and DMFT}

The idea of ``DFT+DMFT'' is to combine the one-particle Hamiltonian
from DFT with an improved description of a selection of states corresponding
to orbitals of limited spatial extension, the ``correlated states'',
within DMFT \citep{LDA+DMFT-anisimov-1997,LDA+DMFT-licht}. While
modern implementations \citep{cRPA-DMFT-LaOFeAs-markus,Wannier-lechermann-2006}
have become increasingly sophisticated, to first approximation, one
can understand DFT+DMFT as the solution of a multi-orbital Hubbard
model, the parameters of which stem from DFT. The DFT+DMFT Hamiltonian
is then written as:

\begin{equation}
H^{DFT+DMFT}=H^{DFT}+H^{int}+H^{DC}
\end{equation}

Here $H^{DFT}$ corresponds to the one-particle Hamiltonian obtained
by Density Functional Theory: 
\begin{equation}
H^{DFT}=\sum_{i,j,\sigma,m}h_{m_{1}m_{2}}^{ij}c_{im_{1}\sigma}^{\dagger}c_{jm_{2}\sigma}
\end{equation}

where $c^{\dagger}$ and $c$ are creation and annihilation operators,
$i$ and $j$ different sites of the crystal, $m$ represent different
orbitals, $\sigma$ is the spin and $h$ is the Hamiltonian.

$H^{int}$ corresponds to the interacting Hamiltonian within the low-energy
model that is used to describe the electronic correlations:

\begin{equation}
H^{int}=\frac{1}{2}\sum_{i}\sum_{(m_{1}\in d,\sigma)\neq(m_{2}\in d,\sigma')}U_{m_{1}\sigma m_{2}\sigma'}n_{im_{1}\sigma}n_{im_{2}\sigma'}
\end{equation}

Here the Hamiltonian includes only density-density interactions for
practical reasons, but it can be generalized to be rotationally invariant.
We note that in practice the full matrix $U_{m_{1}\sigma m_{2}\sigma'}$
can be parametrized using a set of three Slater integrals $F^{0}$,
$F^{2}$, $F^{4}$ and can also depend on the frequency $\omega$
\citep{loig-tesis,Hubbard-interactions}.

Finally, $H^{DC}$ is the double-counting Hamiltonian chosen such
as to avoid double counting of contributions taken into account both
in $H^{DFT}$ and $H^{int}$.

\subsection{Principles of Dynamical Mean Field Theory}

\begin{figure}
\begin{centering}
\includegraphics[width=12cm]{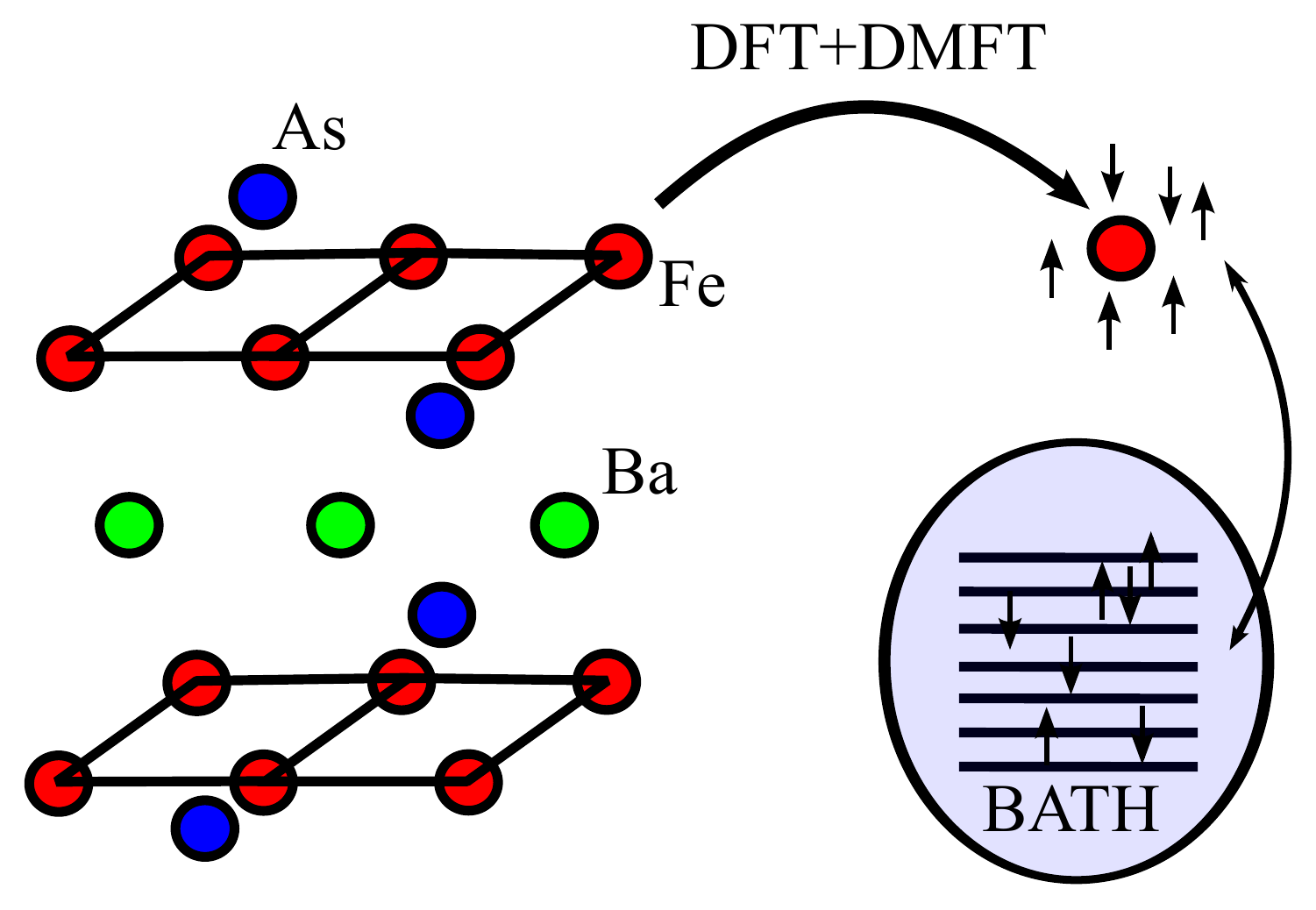} 
\par\end{centering}

\caption[DFT+DMFT mapping of iron pnictides]{Representation of the treatment of a 122 iron pnictide by DFT+DMFT.
The dynamical mean field approximation allows to map the crystal onto
a problem of one atom exchanging electrons with a bath, while the
Wannier orbitals and the free-electron-like hopping probabilities
and electronic energies are calculated from the density functional
theory energy bands.\label{fig:DMFT mapping}}
\end{figure}

In a few words, DMFT consists in replacing the lattice problem by
a single site coupled to a bath, thus making a local approximation
in space. The consequence is that the resulting many-body self-energy
does not have momentum-dependence, but dynamical quantum fluctuations
are fully included: the occupation of the single-site varies as electrons
hop to the bath or from the bath to the site. This process is schematically
represented on Figure \ref{fig:DMFT mapping}.

In practice, a DMFT calculation is a two-step procedure iterated until
self-consistency. The first is the mapping of the lattice model onto
the effective auxiliary quantum impurity problem. The latter is characterized
by the local Hubbard interaction, and a bath propagator describing
hopping between site and bath in the absence of interactions. It can
be thought of as a coupling between the non-interacting bath states
$l$ (of energy $\epsilon_{l}$) and the impurity site by transition
terms $V_{l}$. The creation and annihilation operators on the impurity
are written $c^{\dagger}$ and $c$ respectively, while the corresponding
operators for the bath are written $a^{\dagger}$ and $a$. $n=c^{\dagger}c$
is the number of electrons on the impurity. We can decompose the Hamiltonian
into three parts. One is related to the bath, another to the impurity,
and the third one to the coupling between the bath and the impurity:

\begin{equation}
H_{AIM}=H_{bath}+H_{atom}+H_{coupling}\label{eq: AIM hamiltonian}
\end{equation}

with

\[
H_{atom}=\frac{1}{2}\sum_{(m_{1},\sigma)\neq(m_{2},\sigma')}U_{m_{1}\sigma m_{2}\sigma'}n_{m_{1}\sigma}n_{m_{2}\sigma'}+\sum_{m,\sigma}(\epsilon_{m,\sigma}-\mu)n_{m,\sigma}
\]

\[
H_{coupling}=\sum_{l,m,\sigma}V_{lm\sigma}(a_{l,\sigma}^{\dagger}c_{m,\sigma}+c_{m,\sigma}^{\dagger}a_{l,\sigma})
\]

\[
H_{bath}=\sum_{l,\sigma}\epsilon_{l}a_{l,\sigma}^{\dagger}a_{l,\sigma}
\]

As in any mean field theory, a self-consistency condition is needed
to related the auxiliary problem to the original one. In DMFT, one
imposes that the local Green's function of the lattice, defined as
the sum over the momentum $k$ of the momentum-resolved Green's functions
written in a localized basis set, is equal to the Green's function
of the local model.

Actually, this Green's function of the impurity model only depends
on a hybridization function $\Delta$ which integrates all the degrees
of freedom of the bath:

\[
\Delta_{m}(i\omega_{n})=\sum_{l}\frac{V_{lm}^{2}}{i\omega_{n}-\epsilon_{l}}
\]

through the formalism of an effective action $S_{eff}$:

\[
S_{eff}=-\int_{0}^{\beta}d\tau\int_{0}^{\beta}d\tau'\sum_{m,\sigma}c_{m,\sigma}^{\dagger}(\tau)\mathcal{G}_{0m\sigma}^{-1}(\tau-\tau')c_{m,\sigma}(\tau')+\frac{1}{2}\sum_{(m_{1},\sigma)\neq(m_{2},\sigma')}U_{m_{1}\sigma m_{2}\sigma'}\int_{0}^{\beta}d\tau n_{m_{1},\sigma}(\tau)n_{m_{2},\sigma'}(\tau)
\]

in which the Weiss dynamical mean field $\mathcal{G}_{0}$ is defined
as:

\[
\mathcal{G}_{0}^{-1}(i\omega_{n})=i\omega_{n}+\mu-\epsilon_{0}-\Delta(i\omega_{n})
\]

This means that the impurity model is fully described by the effective
field $\mathcal{G}_{0}$ and the Hubbard $U$.

If we assume that we can solve the model and find the associated Green's
function $G_{imp}$, we then get access to the impurity model self
energy:

\[
\Sigma_{imp}(i\omega_{n})\equiv\mathcal{G}_{0}^{-1}(i\omega_{n})-G_{imp}^{-1}(i\omega_{n})
\]

We now make the DMFT approximation: identifying the impurity model
self-energy with the lattice self-energy,

\begin{equation}
\Sigma(k,i\omega_{n})\simeq\Sigma_{imp}(i\omega_{n})
\end{equation}
 In so doing, we loose the non-local components of the self-energy
(there is no momentum dependence of the self-energy any more) and
approximate the local component by that of the effective local problem.

Consequently, the lattice local Green's function $G_{local}$ is computed
as:

\begin{equation}
G_{local}^{m\sigma}(i\omega_{n})=\int dkG_{local}^{m\sigma}(k,i\omega_{n})=\left(\int dk\left[i\omega_{n}+\mu-\Sigma_{imp}(i\omega_{n})-H(k)\right]^{-1}\right)^{m\sigma}
\end{equation}

The self-consistency condition (the correspondence between the lattice
and the model) requires this new local Green's function to equal the
impurity one ($G_{local}=G_{imp}$), that is, it can be used to update
the dynamical mean field through the Dyson equation:

\[
\mathcal{G}_{0}^{-1}=G_{local}^{-1}+\Sigma
\]

This defines a new local problem that has to be solved, and the cycle
is iterated until the Green's function, Weiss field and self-energy
are converged.

In summary, the Green's function of an electron in the bath is self-consistently
determined, such that the impurity's Green's function is exactly equal
to the local Green's function of the lattice -- within the DMFT approximation
$\Sigma(k,i\omega_{n})\simeq\Sigma(i\omega_{n})$.

The technical challenge is the resolution of the impurity model, which
is computationally demanding but at present is relatively accessible,
notably thanks to the implementation of efficient Monte Carlo algorithms
for Anderson impurity problems \citep{TRIQS-website}.

\subsection{Survival Kit: Properties of the Spectral Function}

In this section, we provide a few useful relations concerning the
one-particle spectral function. For simplicity, we consider the single-orbital
case.

Let us suppose that we have a spectral function $A(k,\omega)$, which
measures the probability of extracting an electron of momentum \textbf{$k$}
and energy $\omega$ \citep{mahan} during the (direct/inverse) photoemission
process. In the case where there is only one energy associated to
each momentum, such as for a free particle, $A(k_{0},\omega)=\delta(\omega(k_{0}))$
($k_{0}$ being fixed). However, in interacting systems, $A(k_{0},\omega)$
displays quasi-particle peaks that have a certain width (corresponding
to the inverse quasi-particle lifetime) and high-energy features can
be generated in addition. Still, by construction $A(k_{0},\omega)\geq0$
and the spectral function is normalized: 
\begin{equation}
\int_{-\infty}^{+\infty}\frac{d\omega}{2\pi}A(k_{0},\omega)=1
\end{equation}
 (Of course, the specific value of the normalization is a choice.
Here we normalize per spin and per orbital. Another possibility would
be to normalize to the total number of states (summing over orbitals
and spin).)

This probability allows us to compute physical quantities, as it codes
the relation between momentum and energy. As a simple example, we
can mention the total number of electrons with momentum $k$:

\begin{equation}
n_{k}=\int_{-\infty}^{+\infty}\frac{d\omega}{2\pi}n_{F}(\omega)A(k,\omega)
\end{equation}

Summing $A(k,\omega)$ over momentum distribution gives the k-integrated
spectral function of the interacting system, that is to say the density
of electrons we can extract from the system by employing a certain
energy, regardless of their momentum. This quantity is thus useful
for comparison to angle-integrated photoemission.

The spectral function can be obtained from the Green's function as:

\begin{equation}
A(k,\omega)=-2ImG_{ret}(k,\omega)
\end{equation}

Inversely, we can also express the Green's functions from the spectral
function via a Hilbert transformation: 
\begin{eqnarray*}
G(k,i\omega) & = & \int_{-\infty}^{+\infty}\frac{d\omega'}{2\pi}\frac{A(k,\omega')}{i\omega-\omega'}\\
G_{ret}(k,\omega) & = & \int_{-\infty}^{+\infty}\frac{d\omega'}{2\pi}\frac{A(k,\omega')}{\omega-\omega'+i0^{+}sign(\omega')}
\end{eqnarray*}

Finally, we can express the spectral function directly in terms of
the self-energy:

\begin{equation}
A(k,\omega)=\frac{-2Im\Sigma_{ret}(k,\omega)}{[\omega-\epsilon_{k}-Re\Sigma_{ret}(k,\omega)]^{2}+[Im\Sigma_{ret}(k,\omega)]^{2}}
\end{equation}

We now turn to a discussion of the link of the spectral function $A$
to ARPES experiments. Indeed, the spectral function can be understood
as an overlap between the ground state with N electrons and excited
states with N-1 or N+1 electrons \citep{mahan}:

\begin{equation}
A(k,\omega)=\begin{cases}
\sum_{e}|<\psi_{e}|c_{k}^{\dagger}|\psi_{0}>|^{2}\delta(\omega+\mu+E_{0}^{(N)}-E_{e}^{(N+1)}), & \omega>0\\
\sum_{e}|<\psi_{e}|c_{k}|\psi_{0}>|^{2}\delta(\omega+\mu-E_{0}^{(N)}+E_{e}^{(N-1)}), & \omega<0
\end{cases}\label{eq:spectral function}
\end{equation}
 with $\mu$ the chemical potential. As shown in Section \ref{sec:ARPES},
the photocurrent can then be written:

\begin{equation}
I(k,\omega)=I_{0}(k,\nu,\overrightarrow{A})A(k,\omega)f(\omega)\ast R(k,\omega)
\end{equation}

$R$ being the resolution function of our ARPES experiment, $f$ the
Fermi function, and $I_{0}$ depending on one-electron matrix elements.

From the Green's function of an interacting system, we can also directly
obtain the Fermi surface of the system, as the locus of the k-points
where the denominator vanishes for $\omega=0$:

\begin{equation}
\mu-\epsilon_{k_{F}}-\Sigma_{m\sigma}(k_{F},0)=0
\end{equation}

Finally, magnetic properties such as susceptibilities, charge-charge
or current-current correlation functions can also be calculated, and
optical conductivities \citep{Tomczak-optics} can be extracted. Thus,
we are able to predict much about the behavior of a solid only by
knowing its Green's function, and we get not only information about
the ground state -- as is the case in DFT -- but also about excitations,
which makes this framework so appealing.

\section{Spectral properties of transition-metal pnictides and chalcogenides}

\label{sec:Spectral properties}

\subsection{General features of the electronic structure of iron pnictides and
chalcogenides}

\begin{figure}
\begin{centering}
\includegraphics[width=14cm]{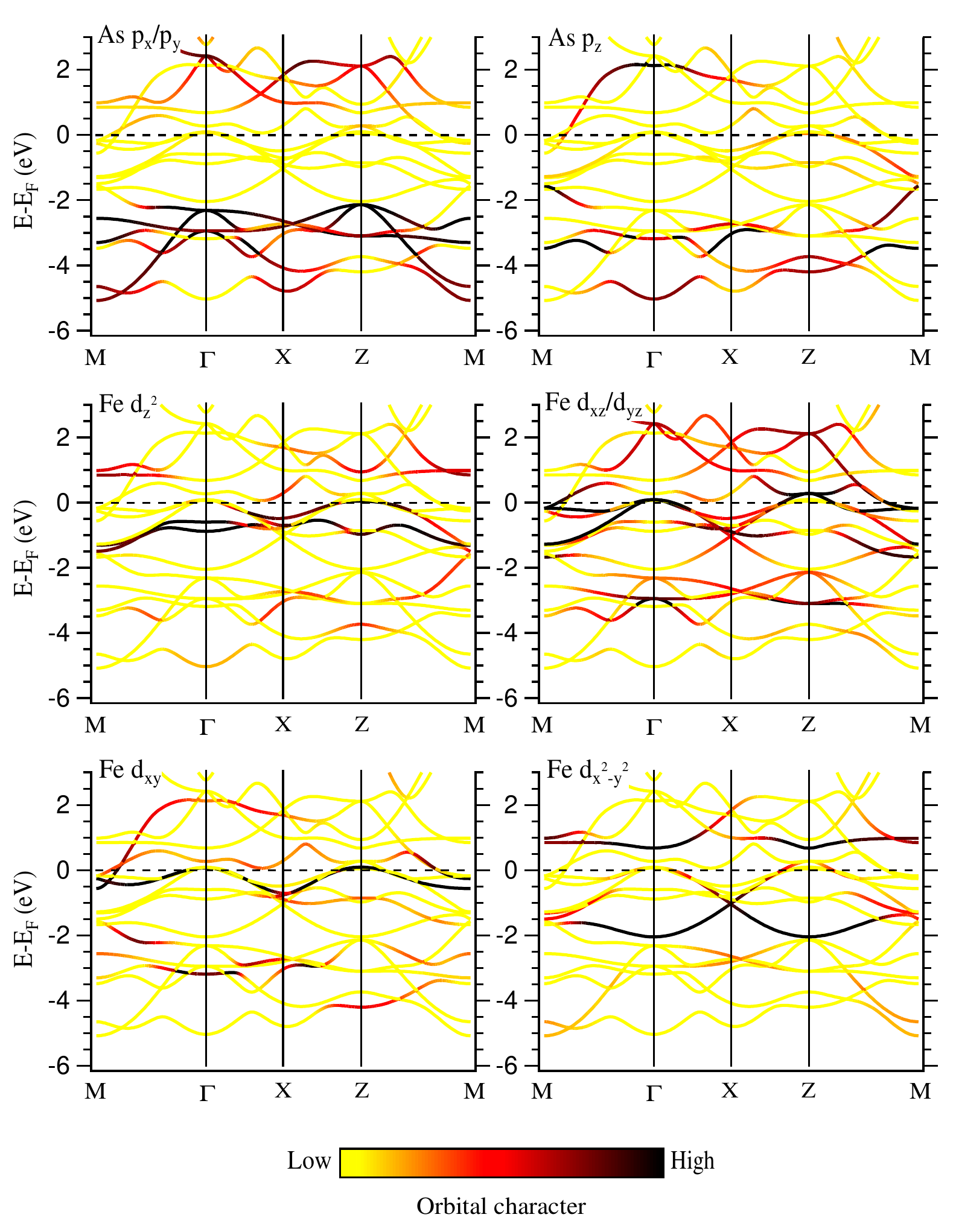} 
\par\end{centering}

\caption[Band Structure of BaFe$_{2}$As$_{2}$]{Kohn-Sham band structure of BaFe$_{2}$As$_{2}$ calculated within
the local density approximation (LDA) to DFT. The experimental lattice
structure (Ref. \citep{Albenque_prb2010}) was used in the calculations.
The color coding indicates the respective orbital character as given
in the upper left corner of each panel.\label{fig:bands}}
\end{figure}

\begin{figure}
\begin{centering}
\includegraphics[width=14cm]{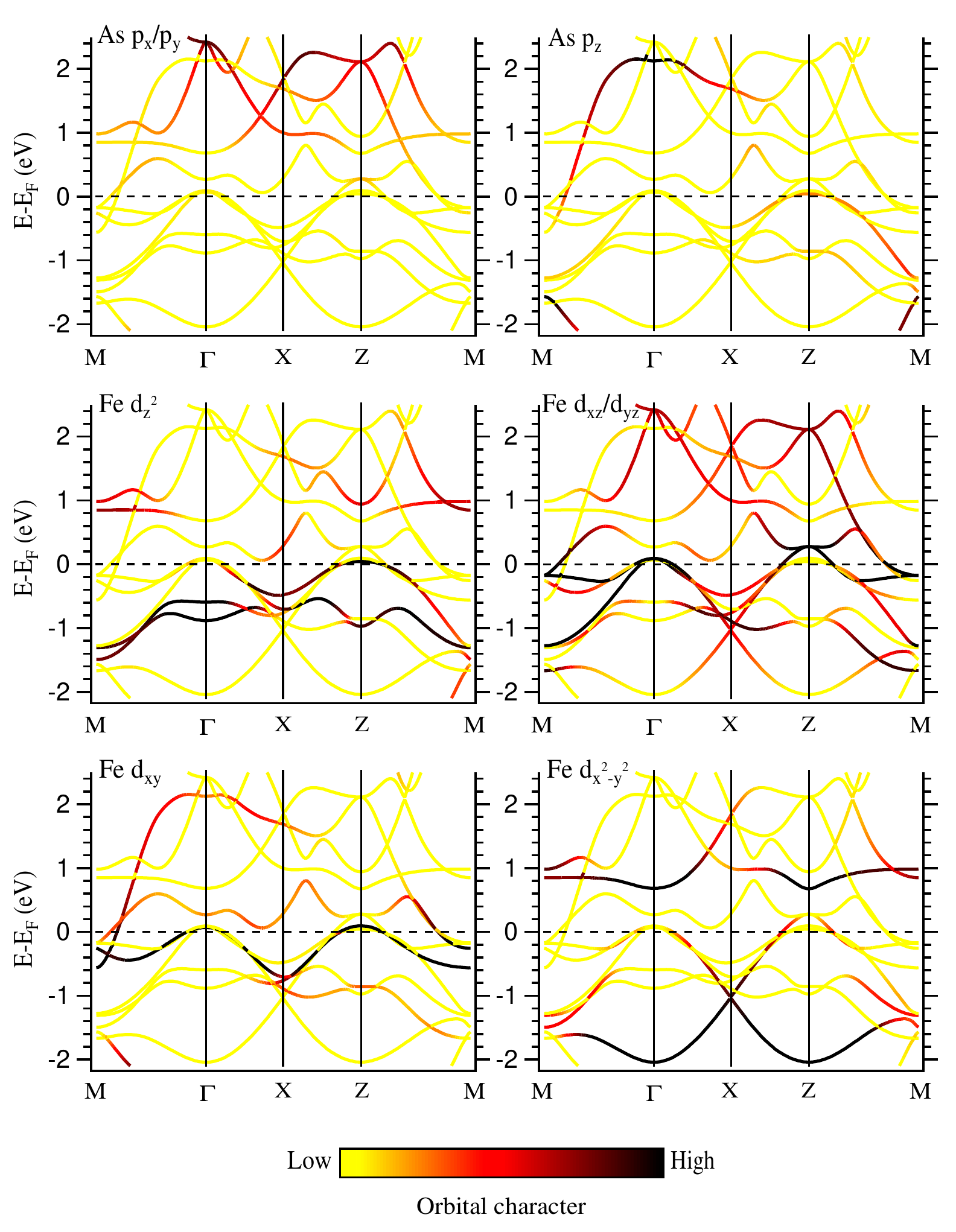} 
\par\end{centering}

\caption[Band Structure of BaFe$_{2}$As$_{2}$]{Same as Fig. \ref{fig:bands} with a zoom on the Fe-\textit{d} bands.\label{fig:bands-zoom}}
\end{figure}

Typical undoped iron pnictides have a nominal band filling of 6 electrons
in the Fe-d shell, and one can draw a phase diagram around this filling
with -- for some materials -- a superconducting dome on each side
-- electron-doped and hole-doped. At this particular filling, realized
in most Fe-based pnictides and chalcogenides (such as pure BaFe$_{2}$As$_{2}$,
LiFeAs, LaFeAsO and FeSe), the Fermi surface is typically formed by
three hole pockets and two electron pockets of Fe-d character at the
$\Gamma$ and $M$ points respectively. Figs.~\ref{fig:bands} and
\ref{fig:bands-zoom} show the DFT-LDA band structure of the prototypical
BaFe$_{2}$As$_{2}$ compound. The iron bands are grouped together
around the Fermi level with a total bandwidth of about 4-5 eV, with
the Fermi level at filling 6/10. The As-p energy bands lie below the
Fe-d bands, approximately between -2 eV and -6 eV. The presence of
an energy gap between iron and arsenic bands depends on the compound,
but quite generally there is non-negligible hybridization between
As and Fe spreading some As-character to the Fe-dominated bands and
vice versa. This effect is visible from the color representation in
Figs.~\ref{fig:bands} and \ref{fig:bands-zoom}. One of the most
interesting points of the one-particle electronic structure of the
iron pnictides is the existence of a ``pseudogap'' in the density
of states of the $d_{x^{2}-y^{2}}$ and $d_{z^{2}}$ orbitals due
to the particular structure of the FeAs plane. It implies that the
weight of these orbitals is generally small in the low-energy excitations
(see Figures \ref{fig:bands} and \ref{fig:bands-zoom}). We stress
that this is an effect of the single particle band structure, not
to be confused with pseudogap behavior induced by many-body correlations.

Substitutions can be performed on every atom of the crystal, which
can result in a wide variety of electronic structures. For example,
substituting iron with cobalt -- 7 electrons in the d shell -- generally
places the compounds close to a ferromagnetic instability due to a
flat band near the Fermi level \citep{BaCo2As2-Nan,Ambroise-BaCo2As2},
while increasing even more the number of electrons will push the Fermi
level close to a back-bending, ``camelback''-shaped energy band
\citep{Nan-Camelback}.

For a detailed analysis of the one-particle band structure of iron
pnictides, we recommend the comprehensive work of Andersen and Boeri
\citep{Andersen-Boeri}.

\subsection{Beyond the single-particle picture: a challenge to theory}

Much theoretical work has been carried out to explore the role of
electronic correlations beyond the band picture, see e.g. \citep{LaOFeAs-haule-2008,cRPA-DMFT-LaOFeAs-markus,cRPA-DMFT-FeSe-markus,Valenti-LiFeAs,LaOFeAs-hansmann-2010,LaOFeAs-anisimov-2009,udyn-werner,Fang-Gutzwiller-pnictides}
for a by far non-exhaustive list of examples. Important early insights
were the discovery of a large dependence of the physics on the Hund's
coupling \citep{LaOFeAs-kotliar-2009, 
Medici-Hund-multiorbital,cRPA-DMFT-FeSe-markus, udyn-werner} and the number of $d$-electrons \citep{liebsch-FeSe, udyn-werner}.
We will come back to these points below.

Early on, the strength of the electronic Coulomb correlations has
appeared to be an important issue but also a source of controversy
\citep{LaOFeAs-kotliar-2009,LaOFeAs-haule-2008,
LaOFeAs-anisimov-2009,LaOFeAs-shorikov,LaOFeAs-anisimov-2008,
cRPA-DMFT-LaOFeAs-markus}, emphasizing the subtleties in the construction of the low-energy
Hamiltonian and the values of the Hubbard and Hund's interactions.
In fact, the resulting spectra of a DMFT calculation depend crucially
on the way the multi-orbital Hubbard-type Hamiltonian is constructed,
concerning the energy window chosen for the low-energy description,
the construction of the corresponding local orbitals, the interaction
terms, and the double counting, see \citep{cRPA-DMFT-LaOFeAs-markus}
for more details. This sensitivity is largely a consequence of the
above mentioned dependence on Hund's coupling and $d$-orbital filling
since different orbital choices for example lead to different definitions
of what one might want to call a $d$-electron. Most importantly,
these difficulties have made iron pnictides and chalcogenides into
a crucial playground for benchmarking different strategies.

Substantial efforts in arriving at a truly \textit{ab-initio} description
were spent for example in \citep{cRPA-DMFT-LaOFeAs-markus}, using
insights of \citep{cRPA-LaOFeAs-miyake}: The approach developed in
that work was based on a Hamiltonian that incorporated both Fe 3\textit{d}
and ligand As and O \textit{p} states as degrees of freedom for LaFeAsO,
but with a Coulomb energy cost on Fe 3\textit{d} orbitals only. The
many-body Hamiltonian was then solved within LDA+DMFT. The effective
interactions for this specific low-energy model were calculated within
the constrained random phase approximation (cRPA) \citep{cRPA-ferdi-2004}
-- an approach for deriving from first-principles the interacting
Hamiltonian within a target subspace that is appropriate for the description
of the low-energy many-body properties. Within this scheme, LaFeAsO
was described as a metal with moderate strength of the electronic
correlations \citep{cRPA-DMFT-LaOFeAs-markus}, whereas larger effects
were found for $\alpha$-FeSe \citep{cRPA-DMFT-FeSe-markus}.

\subsection{ARPES on iron pnictides and chalcogenides -- General remarks}

In the whole pnictide family, photoemission measurements are able
to identify band dispersions and a Fermi surface.

A difficulty stems from the surface sensitivity of ARPES, though.
Indeed, ARPES usually probes the surface in the photon energy considered,
and the measured energies can be quite different from the bulk band
structure. Surface effects were predicted in LaFeAsO in Ref. \citep{Eschrig-LaFeAsO-surface},
due to the polar character of the surface. The most important surface
effects seem to appear in the 1111 family, due to the polar cleavage
surface that creates an important modification of the potential \citep{Pierre-ARPES-review,Yang-LaFeAsO,C_LiuPRB2010}.
Two main effects can appear and can depend on which plane forms the
surface: the appearance of additional bands, which have been probed
for instance in LaFeAsO \citep{Yang-LaFeAsO} or SrFe$_{2}$As$_{2}$
\citep{Hsieh-SrFe2As2}, or the modification of the total electron
count such as in LaFePO \citep{Lu-PhysicaC}. Surface effects have
also been identified in BaFe$_{2-x}$Co$_{x}$As$_{2}$ \citep{vanHeumenPRL106},
as well as in BaCu$_{2}$As$_{2}$ \citep{Pierre-BaCu2As2}, where
a simple structural model suggests a relaxation of the As-position
to be the main effect. In the following, to keep the discussion simple
we will focus on materials without or with limited surface effects.

At first glance, band structures calculated from density functional
theory in the local density approximation (LDA) roughly correspond
to the measured quasiparticles, except that they have to be renormalized
by a factor of about 2 to 5 depending on the material \citep{Pierre-ARPES-review,Yamasaki-FeSe}.
This confirms that the iron pnictides are indeed metals constituted
of itinerant quasiparticles with an enhanced effective mass produced
by intermediate correlations. %
\footnote{It is worth noting that the binding energy of itinerant As-4\textit{p}
states is also underestimated by the LDA \citep{Lu-LaOFeP,Lu-PhysicaC},
though for slightly different reasons: this is the analog of the usual
``band gap'' problem of DFT-LDA.%
} Correlation effects are stronger when moving to the iron chalcogenides:
Indeed, in FeSe, a Hubbard band has been predicted \citep{cRPA-DMFT-FeSe-markus}
and observed \citep{FeSe-PES-Yoshida,Yamasaki-FeSe}.

\subsection{Fermi surfaces: theory vs. experiment}

\begin{figure}
\begin{centering}
\includegraphics[width=9cm]{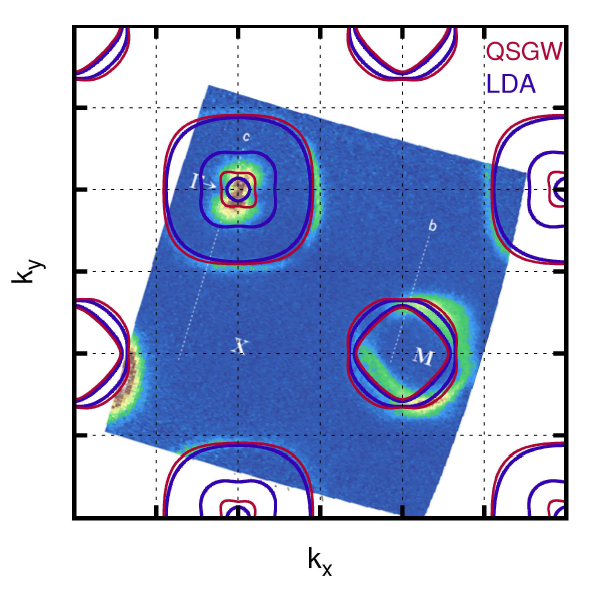} 
\par\end{centering}

\caption[Fermi surface of LiFeAs from ARPES compared to QSGW and LDA]{Fermi surface of LiFeAs from QSGW and LDA compared to the experimental
ARPES spectrum from \citep{Borisenko-LiFeAs}. $k_{z}=0$ plane in
the Brillouin zone for 2 Fe atoms. One of the hole pockets at $\Gamma$
sinks below the Fermi level in GW, and another shrinks drastically,
in agreement with photoemission. (Reprinted with permission from Ref.
\citep{pnictides-QSGW-Jan}, copyright \copyright (2012) by The American
Physical Society.)\label{fig:LiFeAs-FS}}
\end{figure}

The determination of the Fermi surfaces and low-energy excitations
of transition metal pnictides is believed to be a key issue for the
understanding of the phase diagram and mechanism for superconductivity.
ARPES has been used to systematically map out quasi-particle dispersions,
and to identify electron and hole pockets potentially relevant for
low-energy instabilities \citep{ding-EPL-gap,Kaminski-BKFA-FS,Brouet-Nesting,Shin-transition,Golden-ARPES-surface,Fink-BaFe2As2,Malaeb-3d-ARPES}.
Density functional theory (DFT) calculations have complemented the
picture, yielding information about orbital characters \citep{singh-pnictides-bands},
or the dependence of the topology of the Fermi surface on structural
parameters or element substitution \citep{LaOFeAs-veronica,Mazin-Order-Parameter}.

The overall picture emerging from these studies is that for many pnictides
the general topology of the Fermi surface measured within ARPES coincides
with the one obtained from DFT, although quantitative differences
in the pocket sizes (with, most of the time, smaller pockets in nature
than in theory) are common. On those, there seems to emerge some agreement
that combined DFT+DMFT calculations yield corrections in the right
direction, though not always of large enough size. Limitations have
been pointed out, e.g. in Ba(Fe,Co)\textsubscript{2}As\textsubscript{2},
where ARPES has evidenced persisting discrepancies in comparison with
both DFT and DFT+DMFT calculations \citep{FS-Brouet}. In BaFe\textsubscript{2}As\textsubscript{2},
the $d_{xy}$ band is found to form a third hole pocket in the calculations
while a maximum around -150 meV below the Fermi level is captured
by photoemission \citep{Liu-PhysicaC,Vilmercati-3DFS}%
\footnote{It is interesting to note that if the arsenic height is moved away
from the experimental crystal structure in order to minimize the LDA
energy, the position of the $d_{xy}$ band at the $\Gamma$ point
can be dramatically modified \citep{LaOFeAs-veronica,singh-pnictides-bands,pnictides-mazin}.
In some cases, the band structure then found can be closer to the
ARPES measurements, albeit for the wrong reason. In particular, in
BaFe\textsubscript{2}As\textsubscript{2} the maximum around -150
meV captured by photoemission is reproduced \citep{Liu-PhysicaC}.
Within DFT+DMFT, the optimized arsenic height is much closer to the
experimental structure \citep{sc-LDA+DMFT-LaOFeAs-aichhorn,Lee-LiFeAs}.%
}. A similar discrepancy was also revealed in hole-doped Ba\textsubscript{0.6}K\textsubscript{0.4}Fe\textsubscript{2}As\textsubscript{2},
where the band-top of the inner hole pocket, detected from high-temperature
measurements, was found lower than in LDA calculations \citep{Hong-OD-BKFA}.

Even more serious problems arise in FeSe: The LDA Fermi surface takes
the standard form of three hole pockets (of which two appear to be
quasi-degenerate) and two electron pockets. While including correlations
substantially improves on the general band dispersions \citep{cRPA-DMFT-FeSe-markus}
and somewhat reduces the size of the pockets, no drastic changes to
the Fermi surface result. Experimentally, the literature documents
a continuous struggle for determining the number, size and k$_{z}$-dispersion
of the pockets \citep{Lubashevsky_NPhys2012,Maletz-FeSe,Nakayama-FeSe-nematicity,Shimojima-detwinned-FeSe,Okazaki-FeSe-BCS-BEC,Kasahara-FeSe-BCS-BEC,Watson-FeSe-nematic,Zhang-Peng-FeSe-splitting,Suzuki-FeSe-orbital,Zhang-FeSe-nematic},
and the situation is still not fully clear. Still, there seems to
emerge some consensus on \textit{tiny} energy scales of the pocket
depths (measured in meVs and thus of the order of the gap) raising
speculations about highly unconventional types of superconductivity.
Furthermore, rather unconventional behaviors have been observed when
cooling FeSe into its low-temperature orthorhombic phase, where spin-orbit
coupling and possible orbital order are invoked to rationalize the
observed degeneracies and absence thereof. To the best of our knowledge,
no \textit{ab initio} method has so far been able to describe these
properties. For a discussion of the construction of an effective tight-binding
model parameterizing the correlated electronic band structure, see
the Review by P. Hirschfeld in this volume.

Finally, another particularly interesting example is LiFeAs where
DFT-LDA underestimates the size of the outer hole pocket at the $\Gamma$
point, compensated by a gross overestimation of the size of the inner
ones. Experimentally it was difficult to reach agreement even on the
number of pockets since a band just touching the Fermi level is sometimes
but not always resolved as forming a Fermi surface \citep{UmezawaPRL2012,BorisenkoSym2012, Chi, Borisenko-LiFeAs}.

Interestingly, many-body perturbation theory approximating the self-energy
by its first order term in the screened Coulomb interaction W (so-called
``GW approximation'') results in a substantially improved description:
quasi-particle self-consistent (QS)GW \citep{scGW-kotani} calculations
found momentum-dependent corrections to the LDA Fermi surfaces to
result in a suppression of the innermost pocket \citep{pnictides-QSGW-Jan}.
The Fermi surface within QSGW thus shows only two hole pockets around
the center of the Brillouin zone \citep{pnictides-QSGW-Jan} compared
to three in LDA or LDA+DMFT \citep{Singh-LiFeAs,Valenti-LiFeAs,Lee-LiFeAs}
with the overall size in better agreement with photoemission measurements
than the LDA Fermi surface (see Figure \ref{fig:LiFeAs-FS}). These
observations indicate that combined schemes retaining the advantages
of the GW approximation such as the calculation of non-local exchange
and of the non-perturbative computation of correlation effects from
DMFT is a promising way. A similar correction to the Fermi surface
due to non-local exchange was indeed found in the cobalt compound
BaCo$_{2}$As$_{2}$ \citep{Ambroise-BaCo2As2}. We come back to this
point in the section on recent theoretical developments.

\begin{figure}
\begin{centering}
\includegraphics[width=9cm]{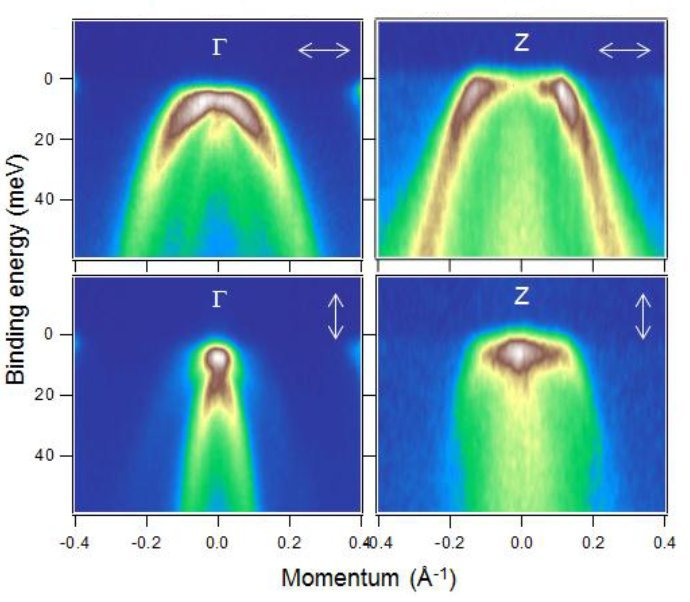} 
\par\end{centering}

\caption[Spin-orbit coupling in LiFeAs]{High-resolution low-temperature ARPES data near $\Gamma$ and Z points
recorded with the light of horizontal (upper) and vertical (lower)
polarizations in LiFeAs. The degeneracy of the two bands at zero momentum
is lifted, possibly by spin-orbit coupling. Reprinted with permission
from Ref. \citep{Borisenko-SOC-LiFeAs}.\label{fig:LiFeAs-SOC}}
\end{figure}

It was argued that the effect of spin-orbit coupling is primordial
for the description of the Fermi surface, since it can lift some degeneracies
very close to the Fermi level in the $d_{xz}/d_{yz}$ orbital space,
notably for the hole pockets at the $\Gamma$ point and for the electron
pockets along the MX direction \citep{Borisenko-SOC-LiFeAs}. The
compounds in which such effects have been discussed are again FeSe
and LiFeAs. Indeed, in LiFeAs two of the usual hole pockets at the
$\Gamma$ point can be distinguished by ARPES, while if the spin-orbit
coupling could be neglected these bands should be degenerate. In this
context, Borisenko and collaborators attempted to extract an estimate
for the spin-orbit coupling strength, arriving at a value of around
10 meV \citep{Borisenko-SOC-LiFeAs} (see Figure \ref{fig:LiFeAs-SOC}).
Nevertheless, some mysteries also remain which might contradict the
interpretation as a pure spin-orbit effect: oddly, the splitting is
temperature dependent; ii) the splitting is strongly doping-dependent;
iii) There is not so much difference between As, Se and Te. Together
with the quite universally observed indications hinting at an influence
of nematic fluctuations in the proximity to structural phase transitions,
one might speculate that the final picture might be a cooperative
interplay of different ingredients. These issues clearly deserve further
investigations.

\subsection{Bandwidth trends and comparison to electronic structure calculations}

\label{sub:bandwidth and LDA}

Assessing the strength of the band structure renormalization allows
us to extract general tendencies concerning the strength of electronic
correlations. Within a same family, and if the crystal structure is
not modified by the introduction of impurities, hole-doping increases
the band renormalization while electron-doping reduces it. This goes
hand in hand with a trend in coherence properties predicted from first
principles calculations \citep{udyn-werner}: while the electron-doped
Ba122 compound is in a Fermi liquid regime, the coherence properties
degrade when moving to the hole-doped side, where Hund's coupling
induces a ``fractional self-energy'' regime at intermediate temperatures,
thus considerably suppressing the coherence temperature where true
Fermi liquid behavior sets in. We will come back to this point below.

The trend on quasi-particle renormalization is most easily seen in
the BaFe\textsubscript{2}As\textsubscript{2} family in the paramagnetic
state: within the same study \citep{Yi-BaFe2As2-family} a factor
of 2.7 was found for the hole-doped Ba\textsubscript{0.6}K\textsubscript{0.4}Fe\textsubscript{2}As\textsubscript{2},
of 1.5 for the undoped BaFe\textsubscript{2}As\textsubscript{2}
and of 1.4 for the electron-doped BaFe\textsubscript{0.88}Co\textsubscript{0.12}As\textsubscript{2}.
The 11 family displays the highest renormalization, with factors found
in the literature larger than 3.5 \citep{Yamasaki-FeSe,FeSe-PES-Yoshida,FeSe-tamai}.
The 111 are an intermediate case with a renormalization factor of
3 -- 4 in LiFeAs and NaFeAs \citep{Borisenko-LiFeAs,He-NaFeAs,Cui-Na(FeCo)As}.
Finally, substituting Fe by Ru in BaFe\textsubscript{2}As\textsubscript{2}
substantially reduces these effects, due to the less localized 4\textit{d}
shell \citep{BaFeAs-brouet,Nan-BaRu2As2}.

\begin{figure}
\begin{centering}
\includegraphics{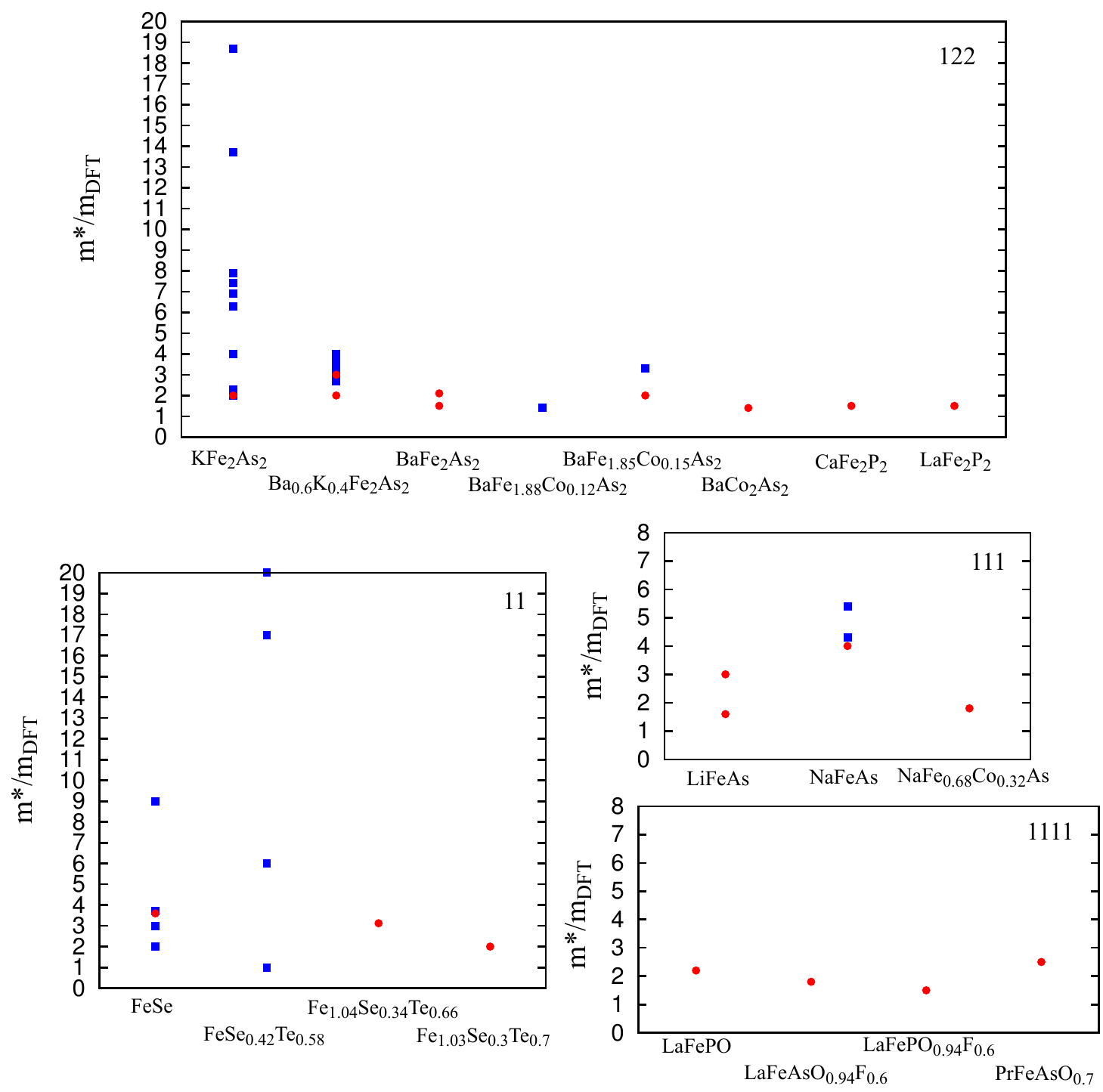}
\par\end{centering}

\caption{Effective mass obtained from ARPES experiments compared to DFT calculations
in different families of iron pnictides. Red dots correspond to an
overall renormalization factor obtained for all bands on an extended
energy range, while blue squares correspond to low-energy, orbital-dependent
or momentum-dependent renormalization factors. Data was taken from
Refs. \citep{Sato-KFe2As2,yoshida-KFe2As2,Hong-OD-BKFA,Yi-BaFe2As2-family,kordyuk-BKFA,Razzoli-pnictides-correlations,Terashima_PNAS2009,brouet-arxiv-Ba(FeCo)2As2,BaCo2As2-Nan}
for the 122, Refs. \citep{Yamasaki-FeSe,Maletz-FeSe,FeSe-tamai,Chen-FeTeSe,Nakayama-FeTeSe}
for the 11, Refs. \citep{Borisenko-LiFeAs,Hajiri-LiFeAs,He-NaFeAs,M_Yi_NJP14,Ye-doping-effects}
for the 111 and Refs. \citep{Lu-LaOFeP,Malaeb-LaFeAsO,Nishi_PRB84}
for the 1111.\label{fig:Effective-mass}}
\end{figure}

Nevertheless, a caveat is in order: Some iron pnictides have been
studied extensively by several groups, such as the BaFe\textsubscript{2}As\textsubscript{2}
family\citep{Fink-BaFe2As2,Malaeb-2008,Feng-orbital-Ba(FeCo)2As2,Hong-OD-BKFA,Brouet-Nesting,Vilmercati-3DFS,Liu-PhysicaC,Sato-KFe2As2,BaCo2As2-Nan,BaCo2As2-Dakha,Fink-Ba(FeCo)2As2},
but the magnitude of the renormalization factors found by different
ARPES groups shows an impressive discrepancy. Figure \ref{fig:Effective-mass}
shows the diversity of renormalization factors obtained by fitting
ARPES data to DFT calculations that can be found in the available
literature. If we compare again with the numbers found by Yi and collaborators
\citep{Yi-BaFe2As2-family}, for electron-doped BaFe\textsubscript{0.84}Co\textsubscript{0.16}As\textsubscript{2}
Brouet and collaborators \citep{brouet-arxiv-Ba(FeCo)2As2} found
a much higher value of 3.3, while for Ba\textsubscript{0.6}K\textsubscript{0.4}Fe\textsubscript{2}As\textsubscript{2}
Ding and collaborators \citep{Hong-OD-BKFA} found a lower overall
factor of 2. Many ARPES authors have signaled additional momentum-dependent
or orbital-dependent shifts of the band structure \citep{Yi-BaFe2As2-family,BaFeAs-brouet,Hong-OD-BKFA,Borisenko-LiFeAs,Fink-BaFe2As2}
or orbital-dependent renormalizations \citep{Lu-LaOFeP,Hong-OD-BKFA,Sato-KFe2As2,Yi-KFe2Se2,FS-Brouet,FeSe-tamai,Feng-orbital-Ba(FeCo)2As2}.
The orbital-dependence of renormalization is clearly captured by DMFT,
and it has even been proposed that iron pnictides are in the proximity
of an orbital-selective Mott transition \citep{Medici-orbital-selective,Medici-Hund-multiorbital,Lanata-orbital-selectivity,Medici-selective-Mott}
(see Section \ref{sub:Weak coupling or strong coupling}). On the
other hand, it is interesting to note that at the GW level, the orbital
renormalizations have been found to be independent of the momentum
at low energy, while momentum-dependent corrections to DFT bands are
due to non-local exchange and correlations \citep{pnictides-QSGW-Jan}.

\subsection{Hund's coupling and ``spin-freezing'' regime}

\label{sub:Coherence-incoherence}

Hund's rule coupling $J_{H}$ plays a fundamental role for the physics
of iron pnictides. Early on, in \citep{LaOFeAs-kotliar-2009}, a coherence-incoherence
crossover in the electronic properties of LaFeAsO was identified,
and the exploration of the -- sometimes counterintuitive -- consequences
of Hund's exchange continues to be an active topic in the field. In
this section, we give some background on how to put different observations
related to this physics into perspective.

As discussed in the preceding sections, band structures calculated
from DFT provide a useful starting point for the description of spectral
properties of iron pnictides, in the sense that most often the overall
picture in terms of Fermi surface topology and quasi-particle dispersions
can be matched (even though, of course, for a quantitative description
shifts or renormalizations may be required). These findings document
the validity of the Landau Fermi liquid picture for these compounds.
Still, the energy scales below which the quasi-particle picture is
well-defined can largely vary -- from compound to compound, under
doping or pressure, and even from orbital to orbital in a given material.

\begin{figure}
\begin{centering}
\includegraphics[width=16cm]{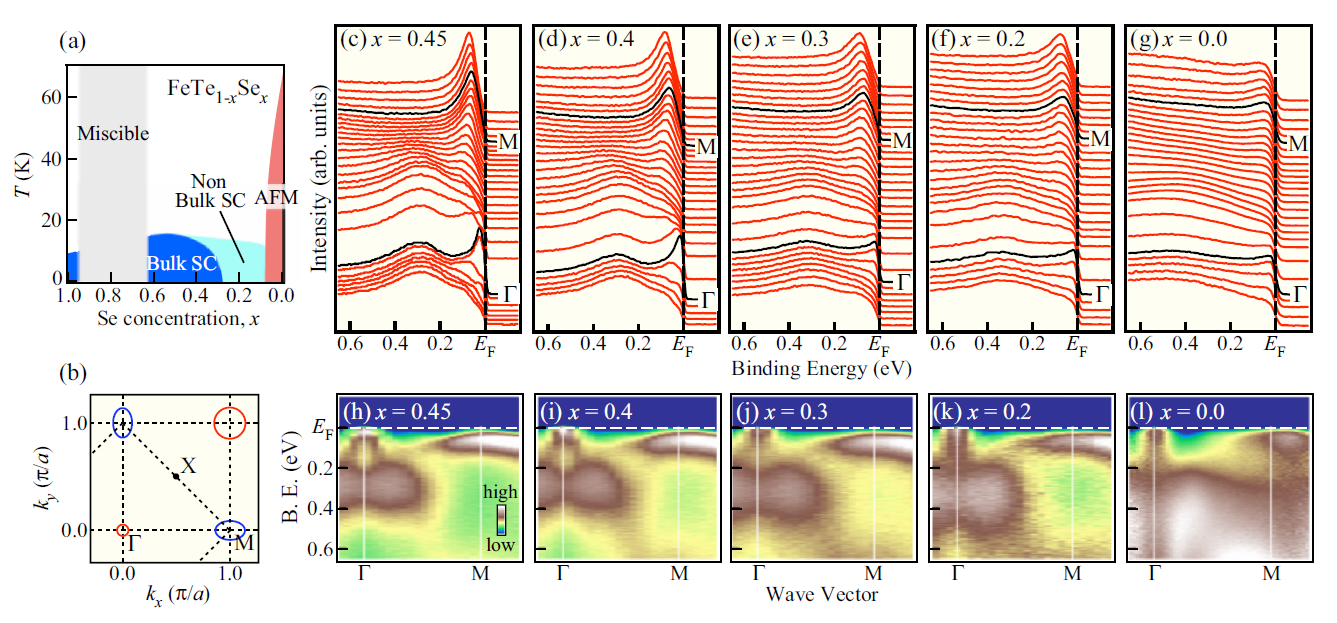} 
\par\end{centering}

\caption[Coherence-incoherence crossover in FeTe\textsubscript{1\textminus{}x}Se\textsubscript{x}]{Coherence-incoherence crossover in FeTe\textsubscript{1\textminus{}x}Se\textsubscript{x }(reprinted
with permission from Ref. \citep{Ieki-Fe(SeTe)}, copyright \copyright
(2014) by The American Physical Society): \protect \protect \protect
\protect \\
 (a) Schematic phase diagram of FeTe\textsubscript{1\textminus{}x}Se\textsubscript{x}
derived from Ref. \citep{Fang-FeSe,Liu-FeSe,Mizuguchi-FeSe}. SC and
AFM denote superconductivity and antiferromagnetism, respectively.
The miscible region exists at high Se concentrations due to the difficulty
in growing single-phase samples.\protect \protect \protect \protect \\
 (b) One-Fe/unit-cell Brillouin zone of FeTe\textsubscript{1\textminus{}x}Se\textsubscript{x}
together with the schematic hole and electron Fermi surfaces at the
and M points, respectively.\protect \protect \protect \protect \\
 (c)\textendash{}(g) Se-concentration dependence of normal-state ARPES
spectra along the -M line in a wide energy region for FeTe\textsubscript{1\textminus{}x}Se\textsubscript{x}
(T = 25 K for x = 0.45\textendash{}0.2 and 80 K for x = 0.0) measured
with the He-I$\alpha$ resonance line ($h\nu$ = 21.218 eV).\protect
\protect \protect \protect \\
 (h)\textendash{}(l) Corresponding ARPES intensity plotted as a function
of binding energy and wave vector.\label{fig:Ieki-crossover}}
\end{figure}

Before coming back to this point, let us note that it is an easy task
to enumerate a whole list of experimental observations that document
effects lying outside the coherent Fermi liquid regime: (a) Ding and
collaborators \citep{Hong-OD-BKFA} noted energy-dependent renormalizations
of the band structure, with an enhanced factor near the Fermi level.
This means that the electronic structure of such compounds cannot
be interpreted in a Fermi-liquid picture with a renormalization factor
$Z$, at least at the temperature of the measurement. (b) Recently,
the existence of a coherence-incoherence crossover as a function of
the isovalent Te-Se substitution has been investigated in Ref. \citep{Ieki-Fe(SeTe)}
(see Figure \ref{fig:Ieki-crossover}). The authors found much shorter
quasiparticle lifetimes in fully substituted FeTe and postulated a
relationship to the development of antiferromagnetism. (c) Finally,
temperature-dependent measurements reveal other surprising effects:
Brouet and collaborators \citep{FS-Brouet} found an evolution of
the number of carriers via k-dependent shifts on the Fermi surface,
while (d) Yi and collaborators \citep{Yi-KFe2Se2} investigated an
orbital-selective coherence-incoherence crossover in alcali-intercalated
iron selenide. This list could be continued nearly \textit{ad libitum}.
We also refer the reader to the detailed review by Florence Rullier-Albenque
in this volume, which deals with transport properties of iron pnictides,
documenting various badly metallic regimes.

\begin{figure}
\begin{centering}
\includegraphics[width=16cm]{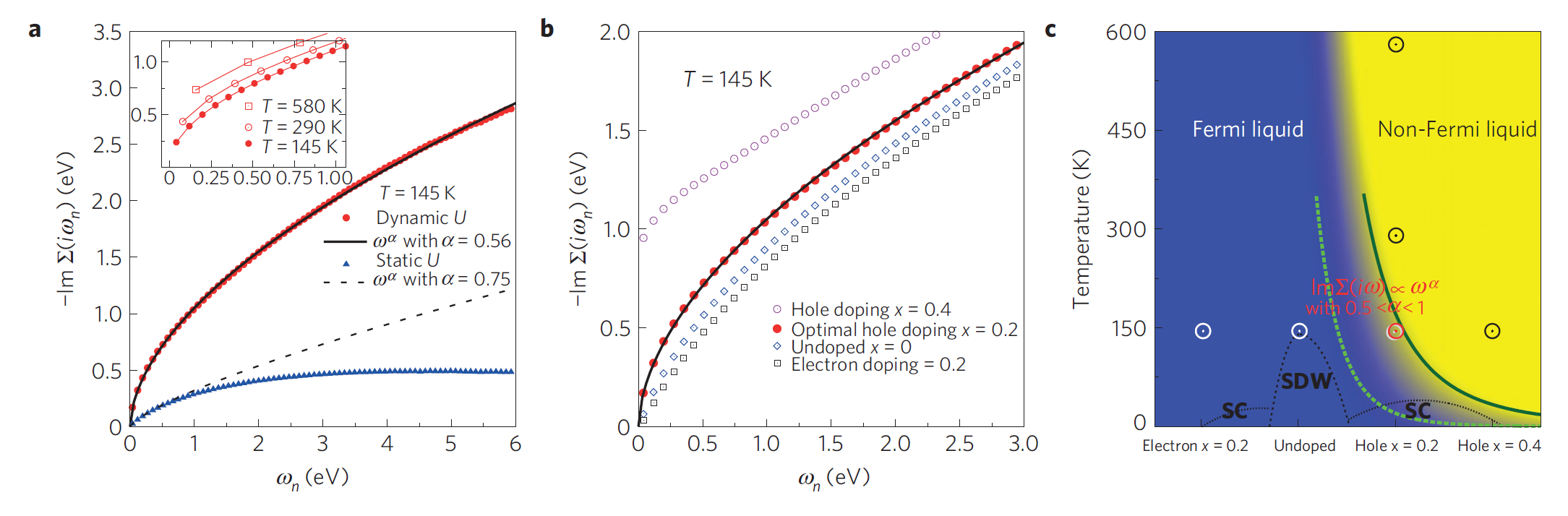} 
\par\end{centering}

\caption[Fractional power-law behavior of the self energy in Ba\textsubscript{1-x}K\textsubscript{x}Fe\textsubscript{2}As\textsubscript{2}]{Fractional power-law behavior of the self-energy in Ba\textsubscript{1-x}K\textsubscript{x}Fe\textsubscript{2}As\textsubscript{2}
(reprinted with permission from Ref. \citep{udyn-werner}).\protect
\protect \protect \protect \\
 (a) Imaginary part of the self-energy (orbital average) on the Matsubara
axis for optimal hole-doping (x=0.2 per Fe) within LDA+DMFT, with
dynamical interaction $U(\omega)$ (red circles) and static $U(\omega=0)$
(blue triangles). Solid and dashed lines are fits of the low-frequency
behavior of the function $-\text{Im}\Sigma(i\omega_{n})=A(\omega_{n}){}^{\alpha}$.
The inset shows the low-frequency behavior of the dynamic-$U$ result
for different temperatures. As the temperature is raised, the extrapolation
$\omega_{n}\longrightarrow0$ yields a non-zero intercept, which indicates
that even excitations at the Fermi level exhibit a finite lifetime.\protect
\protect \protect \protect \\
 (b) Low-energy behavior of the self-energy as a function of doping.
Fermi-liquid behavior is found in the undoped and electron-doped compounds,
whereas a non-zero intercept appears in the overdoped case.\protect
\protect \protect \protect \\
 (c) Sketch of the phase diagram in the space of temperature and doping.
The blue region indicates Fermi-liquid behavior, whereas yellow indicates
a frequency dependence of the self-energy that is not compatible with
Fermi-liquid theory. The light green dashed line marks the boundary
of the crossover region, where the exponent $\alpha$ starts to deviate
from 1. The dark green solid line corresponds to $\alpha=0.5$, which
marks the ``spin-freezing'' transition. To the right of this line,
an incoherent metal phase with a non-zero intercept of $\text{Im}\Sigma$
is found. The experimentally measured phase diagram with superconducting
(SC) and spin-density wave (SDW) ordered phases is indicated by black
dotted lines. Full substitution (KFe\textsubscript{2}As\textsubscript{2})
corresponds to x=0.5.\label{fig:Werner}}
\end{figure}

On the theoretical side, we highlight a theoretical study that was
performed independently of the discovery of superconductivity in the
iron pnictides: a DMFT study of a three-band orbital Hubbard model
with a local interaction parameterized by Hubbard interactions and
Hund's exchange at filling 2/6 or 4/6 \citep{werner-spin-freezing}
revealed the existence of a ``spin-freezing'' regime characterized
by a power-law behavior of the imaginary part of the self-energy in
Matsubara frequencies, corresponding to an incoherent metal. This
regime is characterized by strong Hund's coupling favoring a high-spin
state strongly suppressing charge fluctuations. Kondo screening of
the effective impurity in the DMFT description -- which, in the DMFT
language, is a necessary condition for the formation of a coherent
Fermi liquid state -- is thus suppressed and Fermi liquid behavior
is recovered only below a possibly extremely low coherence temperature.
Most interestingly, in \citep{udyn-werner}, such a power-law regime
of the self-energy was later on detected in calculations for iron
pnictides, namely for hole-doped BaFe$_{2}$As$_{2}$ (see Figure
\ref{fig:Werner}). While in a three-orbital model, the spin-freezing
regime is realized around fillings 2/3 or 4/6, that is, around even
commensurate fillings, the analogous situation in a 5-orbital system
includes precisely the 6/10 filling of the nominal valence of the
iron pnictides. In \citep{udyn-werner}, the phase diagram around
this filling -- corresponding to pure, hole- or electron-doped BaFe$_{2}$As$_{2}$
was found to display the same phenomenology as in the three-orbital
model: Fermi liquid behavior was enhanced (that is, the coherence
temperature increased) under electron doping, moving the system further
away from the half-filled 5/10 case, while the incoherent ``spin-freezing''
regime was evidenced as the intermediate temperature regime under
hole-doping, that is, for fillings (6-x)/10. The exploration of the
consequences of Hund's coupling in iron pnictides has therefore become
an active area of research, including the limits of the local description
of DMFT \citep{Nomura-Hund-nonlocal} or analytical insights from
a renormalization group treatment \citep{Aron-Hund-RG}. As discussed
on the basis of zero-temperature calculations \citep{Medici-selective-Mott},
at low-temperatures (that is, within the coherent regime) the filling
dependence of the Fermi liquid properties translates into an evolution
from a more strongly correlated state (characterized by a small quasi-particle
weight Z) close to half-filling to a less correlated state at fillings
6+x.

Model studies have pushed this picture further by analyzing the orbital
dependence of the renormalization factors: De' Medici and collaborators
have argued that the strong Hund's coupling can decouple the Fe-\textit{d}
orbitals and drive them close to a Mott-selective phase \citep{Medici-selective-Mott,Medici-orbital-selective,Medici-Hund-multiorbital,Lanata-orbital-selectivity}
where the $d_{xy}$ orbital would be the most strongly correlated
orbital. While this general trend agrees well with results from more
realistic DMFT calculations (in particular, concerning the less correlated
$z^{2}$ and $x^{2}-y^{2}$ orbitals, due to their one-particle pseudogap
\citep{cRPA-DMFT-FeSe-markus}), the quantitative aspects of this
differentiation seem to be considerably stronger in the model context
where only the five $d$-orbitals are retained, than in electronic
structure calculations where also pnictogen/chalcogen $p$-states
are included.

\subsection{Weak coupling or strong coupling? -- Are iron pnictides siblings
of cuprates?}

\label{sub:Weak coupling or strong coupling}

As we have seen, in the pnictides and chalcogenides the variety of
compounds and fillings of the $d$ shell combined to the great sensitivity
of correlations due to the large Hund's coupling account for a wide
diversity of behaviors. In this section, we will focus on the more
strongly correlated features and on the possible similarities with
the cuprates. As alluded to at several points in this review already,
the most correlated family are the chalcogenides, notably due to the
smaller hybridization between Fe and Se. In FeSe, the existence of
a broad feature of Fe-\textit{d} character around -2 eV in the photoemission
spectrum has been revealed in Ref. \citep{FeSe-PES-Yoshida,Yamasaki-FeSe}
and was identified as a Hubbard band by Aichhorn and collaborators
\citep{cRPA-DMFT-FeSe-markus}. The stronger correlations in FeSe
also come along with a more pronounced orbital differentiation, namely
more strongly correlated $xy$- and $xz/yz$-orbitals. This is quite
natural due to the (single-particle-) pseudogap of the $z^{2}$ and
$x^{2}-y^{2}$ orbitals at the Fermi level \citep{cRPA-DMFT-FeSe-markus}.

In Fe\textsubscript{1.06}Te, photoemission found a pseudogap in the
paramagnetic phase, progressively closing when the temperature is
reduced in the antiferromagnetic phase \citep{Lin-FeTe-pseudogap}.
Interestingly, a pseudogap was also measured in underdoped Ba\textsubscript{1-x}K\textsubscript{x}Fe\textsubscript{2}As\textsubscript{2}
\citep{Xu-BKFA-pseudogap}, and the relation between those two experiments
remains to be understood. More experimental evidence would be desirable
to confirm the existence of a pseudogap in iron pnictides and probe
its characteristics.

\begin{figure}
\begin{centering}
\includegraphics[width=12cm]{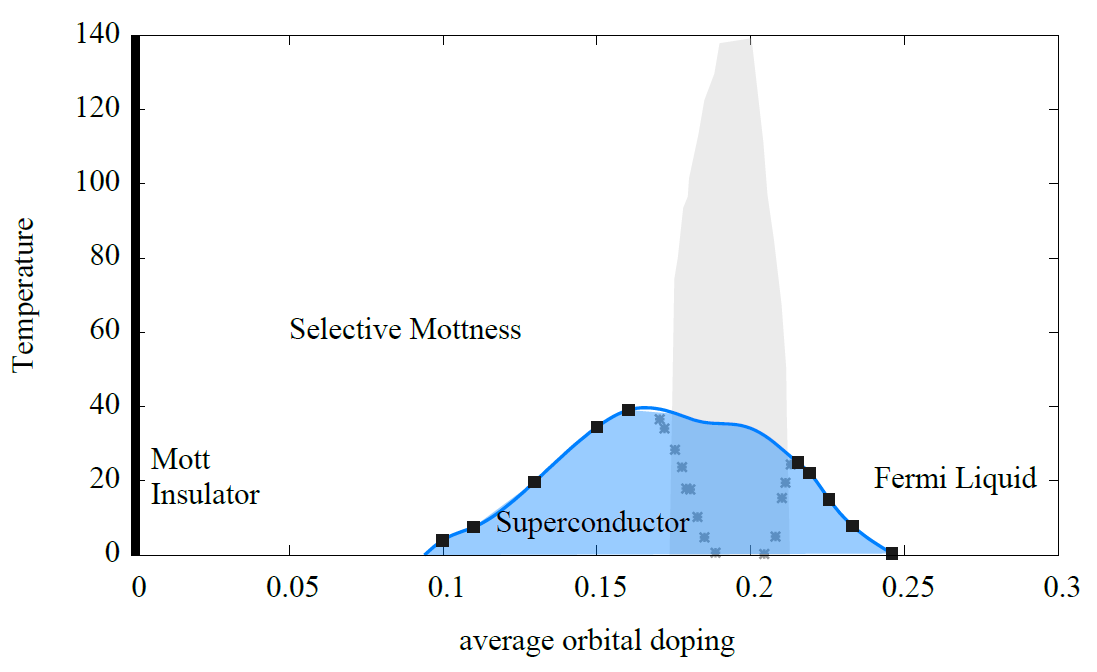} 
\par\end{centering}

\caption[Possible unified phase diagram]{Possible unified phase diagram for pnictides and cuprates reprinted
with permission from Ref. \citep{Medici-selective-Mott}, copyright
\copyright (2014) by The American Physical Society. Plotted here
is the experimental phase diagram for doped BaFe$_{2}$As$_{2}$ as
a function of the average orbital doping. In the hypothesis that the
magnetically ordered/orthorombically distorted phase (grey area) is
accidentally favored by the total commensurate filling n=6 around
the stoichiometric compound thus suppressing the superconductivity
of the tetragonal phase, de' Medici and collaborators artificially
eliminate it and complete the superconducting dome (blue area).\label{fig:Cuprate-pnictide}}
\end{figure}

Finally, we mention that in the quest of similarities with the cuprates,
several authors have argued in favor of an analogy of the phase diagrams
when the average filling per orbital is considered \citep{Medici-selective-Mott,Medici-orbital-selective,Medici-Hund-multiorbital,Lanata-orbital-selectivity,
Misawa}. In this picture, the behavior at filling $d^{6}$ is considered
as a (magnetic) accident artificially separating the electron- and
hole-doped superconducting regions in the phase diagram, and the proximity
to the Mott insulator (at filling $d^{5}$) would be the key feature
linking cuprate and pnictide physics. The corresponding tentative
phase diagram is shown in Figure \ref{fig:Cuprate-pnictide}.

\subsection{Spectral properties in a nutshell}

Summarizing, at present, there is a large consensus on the fact that
iron pnictides display intermediate correlations -- stronger in the
11 family and chalcogenides \citep{cRPA-DMFT-FeSe-markus}, weaker
in the 1111 and phosphides in general -- driven by the physics of
strong Hund's coupling in multiorbital systems \citep{hund-medici,werner-spin-freezing}.
The consequence is that these materials tend to display a bad-metal
behavior, in which the effective mass and the coherence properties
are very sensitive to the effective number of electrons in the correlated
shell \citep{udyn-werner,BaCo2As2-Nan}, which can be tuned by small
details of the electronic structure, notably the hybridization between
the metal and the pnictogen \citep{Fang-Gutzwiller-pnictides,Ambroise-CaFe2As2}.

\section{Superconducting gap}

\label{sec:SC gap}

\begin{figure}
\begin{centering}
\includegraphics[width=16cm]{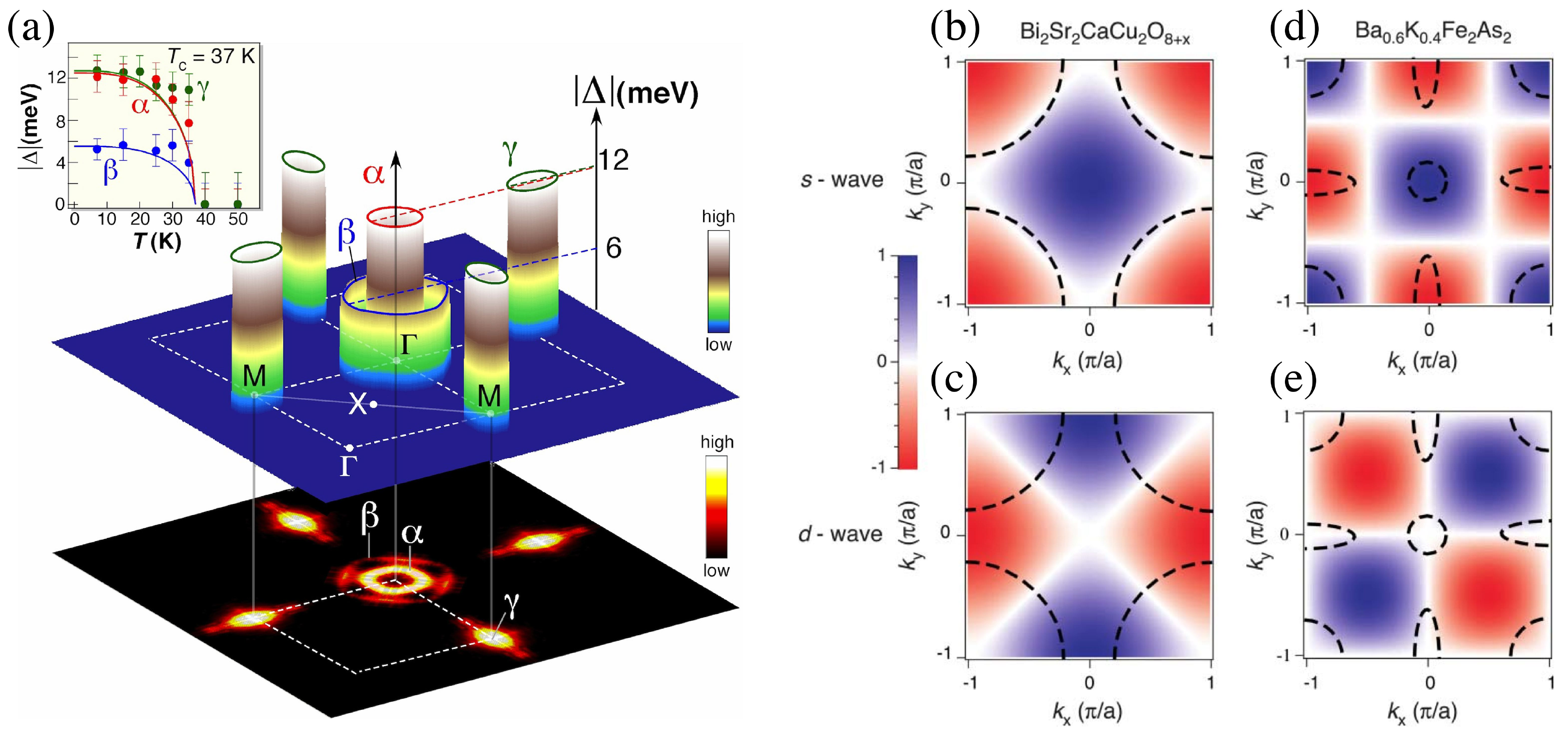} 
\par\end{centering}

\caption{\label{Fig_SC}(a) Three-dimensional plot of the SC gap size ($\Delta$)
in Ba$_{0.6}$K$_{0.4}$Fe$_{2}$As$_{2}$ measured at 15 K on three
Fermi surface sheets (shown at the bottom as an intensity plot) and
their temperature evolutions (inset). Reprinted with permission from
\citep{Ding_EPL}, copyright \copyright\ (2008) by the European
Physical Society. (b)-(e) Visualization of the overlap between Fermi
surfaces and gap functions: (b) $s$-wave $\cos(k_{x})+\cos(k_{y})$
for optimally doped cuprate Bi$_{2}$Sr$_{2}$CaCu$_{2}$O$_{8+x}$;
(c) $d$-wave $\cos(k_{x})-\cos(k_{y})$ for Bi$_{2}$Sr$_{2}$CaCu$_{2}$O$_{8+x}$.
(d) $s$-wave $\cos(k_{x})\cos(k_{y})$ for optimally doped ferropnictide
Ba$_{0.6}$K$_{0.4}$Fe$_{2}$As$_{2}$. (e) $d$-wave $\sin(k_{x})\sin(k_{y})$
for Ba$_{0.6}$K$_{0.4}$Fe$_{2}$As$_{2}$. The color bar indicates
the values of the SC order parameters. Panels (b)-(e) are reprinted
by permission from Macmillan Publishers Ltd: Sci. Rep. \citep{HuJP_SR2012},
copyright \copyright\ (2012).}
\end{figure}

The previous section raised an important question that is relevant
not only to the electronic band structure itself, but to the pairing
mechanism as well: weak or strong coupling? To discuss this question,
we discuss the order parameter of the superconducting transition,
which consists in a complex function of the momentum whose amplitude
corresponds to the size of an electronic band gap $2\Delta$ developing
symmetrically with respect to the Fermi level while cooling down a
sample below its superconducting transition temperature. This amplitude
$\Delta$ is accessible by ARPES directly in the momentum space \citep{Ding_EPL,L_Zhao,Kondo-NdFeAsO,Nakayama-FeTeSe,Borisenko-LiFeAs,Liu-NaFeCoAs}).
The Fermi surface of the first Fe-based superconductors studied \citep{Ding_EPL,L_Zhao}
contributed to popularize a simple pairing mechanism called quasi-nesting
\citep{Mazin-gap, MazinPhysicaC2009, Kuroki,
Graser_NJP2009,Richard_JPCMreview}. Indeed, as shown in the bottom part of Fig.~\ref{Fig_SC}(a) and
discussed the previous section, the Fermi surface is formed by hole
pockets centered at $\Gamma$ and electron pockets centered at M.
Within the quasi-nesting model, the pairing amplitude is largely enhanced
by electron-hole inter-pocket scattering with the antiferromagnetic
wave vector, which coincides with the wave vector from $\Gamma$ to
M. Accordingly, the top of Fig.~\ref{Fig_SC}(a) indicates that a
large gap is observed on every Fermi surface pockets except the $\beta$
band, which is much larger than the others and thus not well quasi-nested
with any other band. For this mechanism to be effective, it is necessary
that the pockets have similar sizes and shapes, but not equal since
this would favor a density-wave ground state \citep{Cvetkovic_EPL2009}.
Therefore, this weak coupling mechanism is by nature very sensitive
to the Fermi surface topology.

There is now some evidence against the weak coupling approach as driving
force of the pairing of electrons in the Fe-based superconductors
\citep{Richard_JPCMreview}. Perhaps the most significant one is the
existence, for the same 122 crystal structure, of at least three distinct
Fermi surface topologies leading to superconductivity, which is the
equivalent of three fundamentally different sets of critical parameters,
a possibility not appealing physically. Indeed, Ba$_{1-x}$,K$_{x}$Fe$_{2}$As$_{2}$
has the Fermi surface topology displayed in Fig. \ref{Fig_SC}(a)
and refined later \citep{Nakayama_EPL2009}. With approximately the
same critical temperature as Ba$_{1-x}$,K$_{x}$Fe$_{2}$As$_{2}$,
the 122-ferrochalcogenides A$_{x}$Fe$_{2-x}$Se$_{2}$ \citep{JG_Guo_PRB2010,FangMH_EPL2011}
have a drastically different Fermi surface topology with no hole pocket
\citep{Qian_PRL2011,XP_WangEPL2011,D_MouPRL2011,Y_Zhang_NatureMat2011,XP_WangEPL2012}.
Finally, due to a chemical potential shift resulting from hole-doping,
the electron pockets at M are replaced by four petal-shaped lobes
in KFe$_{2}$As$_{2}$, which are off-M centered \citep{Sato_PRL2009,Yoshida_JCPS72}.

In contrast to the weak coupling approach, a strong coupling approach
to the pairing is much more robust to the evolution of the Fermi surface.
This is the case of the $J_{1}-J_{2}-J_{3}$ model, where $J_{1}$,
$J_{2}$ and $J_{3}$ represent the exchange parameters between first,
second and third nearest neighbor, respectively. The philosophy of
this approach is to parameterize the magnetic structure of a parent
compound by effective coupling parameters $J_{i}$. We note that --
as shown first by Yaresko et al. \citep{Yaresko-magnetic-FS} -- a
more appropriate parameterization is provided by assuming biquadratic
exchange terms. For the sake of simplicity, we stick here to the simpler
$J_{1}-J_{2}-J_{3}$ model, but refer the reader to the detailed discussion
of the biquadratic model in the Review by P. Hirschfeld in this volume.
In momentum space, the structure of the interactions determines a
form factor consisting of combinations of sine and cosine functions.
If the pairing is induced by the fluctuations of the local moments
away from long-range magnetic ordering, the pairing symmetry can be
assumed to have the same form factor. However, a few possibilities
exist due to the relative phase of the different interactions. Given
the fact that the energy of the whole system should be lowered by
gapping electronic states near the Fermi level, one can assume the
favored symmetry to be the one which maximizes the overlap between
the Fermi surface and the antiferromagnetic form factor \citep{HuJP_SR2012,Richard_JPCMreview}.
We first illustrate this with the cuprates. These materials are well
known for having a dominant $J_{1}$ antiferromagnetic exchange parameter
which leads to a possible $s$-wave gap function and a possible $d$-wave
gap function, as shown in Figs. \ref{Fig_SC}(b) and \ref{Fig_SC}(c),
respectively. Despite nodes, the overlap with the $d$-wave gap function
is much better. In contrast, the dominant antiferromagnetic exchange
parameter in the Fe-ferropnictides is $J_{2}$, which also leads to
different $s$-wave and $d$-wave gap functions (see Figs. \ref{Fig_SC}(d)
and \ref{Fig_SC}(e)), and in that case the overlap of the Fermi surface
is better with the $s$-wave (the so-called $s_{\pm}$ gap function),
with the same gap amplitude for the $\Gamma$-centered hole pockets
and M-centered electron pocket that have the same size, as observed
experimentally. Interestingly, in the ferrochalcogenides neutron data
indicate a non-negligible $J_{3}$ in addition to $J_{2}$, which
induces an asymmetry of the gap function between the $\Gamma$ and
M points. Such an asymmetry would be consistent with what has been
found experimentally for FeTe$_{0.55}$Se$_{0.45}$ \citep{H_Miao_PRB2012}.

Still, we stress again the simplistic nature of these fully localized
models, let it be the $J_{1}-J_{2}-J_{3}$ model or its biquadratic
refinements. On conceptual grounds it is difficult to justify a purely
strong coupling approach. In practice, a much debated case is LiFeAs,
where a measured noticeable anisotropy in the superconducting gap
around the M point \citep{UmezawaPRL2012,BorisenkoSym2012} has been
interpreted as evidence for effects beyond the localized picture.
An alternative scenario, dubbed orbital anti-phase $s_{\pm}$ gap
function, has been proposed on theoretical grounds \citep{Yin_ZP_orbital2014, X_LuPRB85}.
This is a variant of the $s_{\pm}$ scenario, where the phase of the
superconducting gap depends on the orbital character. The observation
by ARPES of an in-gap impurity states in Ba$_{1-x}$,K$_{x}$Fe$_{2}$As$_{2}$
was argued to be consistent with this hypothesis \citep{P_ZhangPRX2014}.

\section{Recent theoretical developments}

A topic that has raised much interest recently is the \textit{ab initio}
determination of the Hubbard interactions, including their dynamical
character and the consequences thereof. The impact of the energy-dependence
of the Hubbard interactions -- $U(\omega)$ as calculated within constrained-RPA
-- on the low-energy properties and the coupling between electronic
and plasmonic excitations in the many-body calculations, has been
investigated in prototypical models \citep{udyn-michele} and then
applied to BaFe$_{2}$As$_{2}$ \citep{udyn-werner} in a study revealing
the spin-freezing phase in the hole-doped side of the phase diagram
(see Figure \ref{fig:Werner}). The inclusion of energy-dependent
Hubbard interactions within a dynamical version of LDA+DMFT \citep{udyn-michele,udyn-werner},
leads to a reduction of the quasi-particle weight at the Fermi energy,
compared to the standard LDA+DMFT with a static interaction. The spectral
weight is shifted to additional satellites at larger energies corresponding
to plasmon excitations. This transfer of spectral weight is accompanied
by a corresponding reduction of the bandwidth, and can be understood
as an effective renormalization of the one-particle hopping matrix
elements by the dynamical screening processes. The quantitative aspects
of this effect depend on the spectrum of (bosonic) screening processes
as encoded in the frequency-dependence of $U(\omega)$ and have been
worked out in \citep{udyneff-michele}. The values of the reduction
factor Z$_{B}$ are between 0.59 and 0.63 for LaFeAsO, FeSe and BaFe$_{2}$As$_{2}$
\citep{udyneff-michele}, which suggests a further enhancement of
the effective masses computed by static LDA+DMFT by a factor of at
least 1.6 -- 1.7 in iron pnictides and raises again the controversy
about the precise strength of correlations.

\begin{figure}[!tph]
\centering{}\includegraphics[width=9cm]{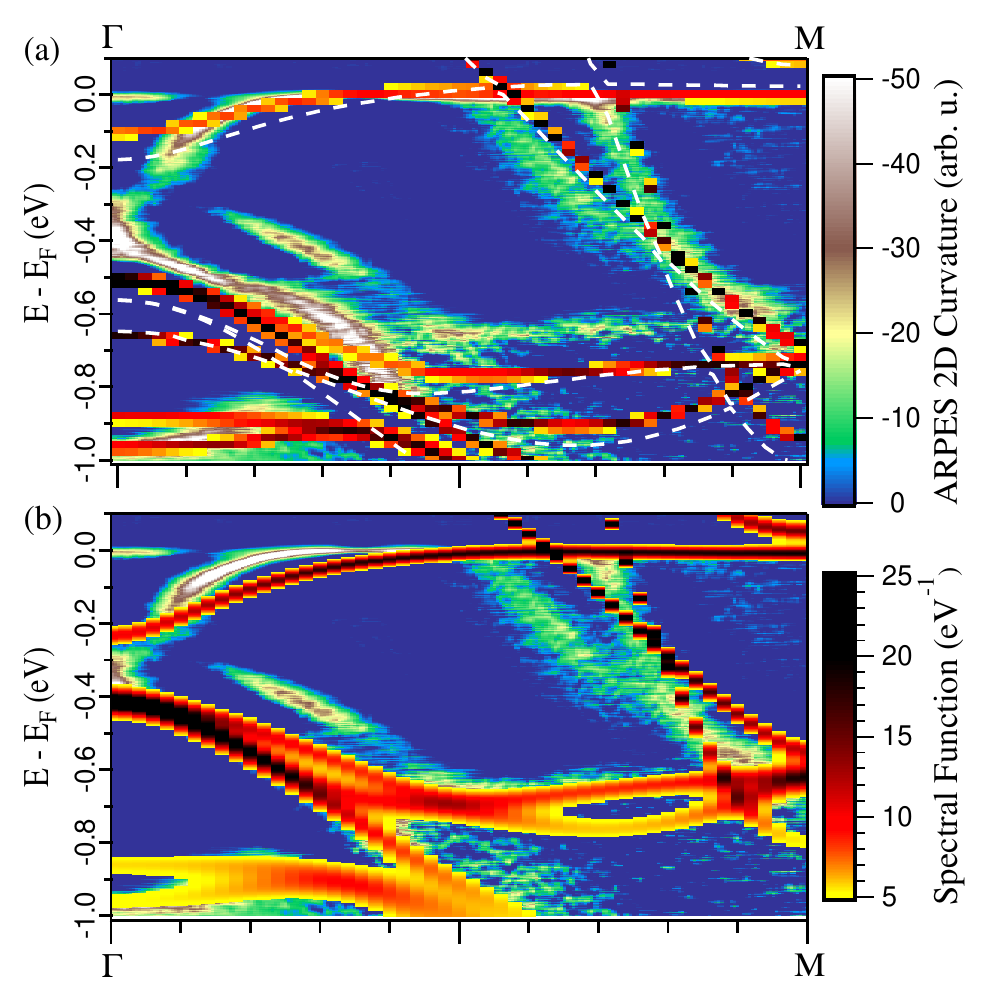}
\caption{Bands of BaCo$_{2}$As$_{2}$ along the $\Gamma$M direction, extracted
from the spectral function calculated by (a) SEx+DDMFT and (b) LDA+DMFT,
superimposed on ARPES data from Ref.~\citep{BaCo2As2-Nan} (represented
as a second derivative of the photoemission intensity). The Quasiparticle
Self-consistent \textit{GW} band structure is also given (white dashed
lines). Reprinted with permission from Ref. \citep{Ambroise-BaCo2As2},
copyright \copyright (2014) by The American Physical Society.\label{fig:BaCo2As2}}
\end{figure}

As discussed above, combined DFT+DMFT has been very successful in
describing spectral properties of transition metal pnictides, concerning
quasi-particle renormalizations, onset of incoherent behavior or even
two-particle quantities. Still, quite generally, problems remain concerning
the size of the pockets at the Fermi energy: it seems that DFT+DMFT
just as the Kohn-Sham band structure of DFT overestimates the size
of Fermi surface pockets. Since in such calculations the overall particle
number fixes the chemical potential self-consistently, an overestimation
of hole pockets is necessarily accompanied by an overestimation of
electron pockets and vice versa. The additional correction needed
is a correction that would be able to ``shift the pockets apart'',
that is shift hole pockets downwards and electron pocket upwards.
In cases, where the pockets are built from different orbital contributions,
this would be possible by a purely local but orbital-dependent self-energy
correction as provided by local approximation such as DMFT. In the
pnictides, however, the requirement is even stricter: xz/yz orbitals
form both electron and hole pockets, and shifting those apart indicates
additional k-dependent (that is, nonlocal) corrections. The combined
$GW$+DMFT scheme \citep{GW+DMFT-biermann,SrVO3-GW+DMFT-Jan,SrVO3-GW+DMFT-Jan-long,Thomas-PRL-GW+DMFT,GW+DMFT_Thomas_PRB}
is a promising way to address this challenge. Recently, an even cheaper
way has been proposed, namely a scheme that incorporates dynamical
screening and non-local exchange into a DMFT calculation \citep{Ambroise-BaCo2As2,Ambroise-SrVO3, Ambroise-CaFe2As2}.
Interestingly, this technique also reconciles the Stoner criterion
with the experimental absence of magnetism in BaCo$_{2}$As$_{2}$.
We finish this review by a brief description of these very recent
developments.

At the pure $GW$ level, it has been found that the self-energy can
be roughly separated into a non-local static part and a local -- in
the sense that in can be expressed in a local basis of Wannier functions
-- dynamical part at low-energy in the iron pnictides \citep{pnictides-QSGW-Jan}
and in the transition metal oxide SrVO\textsubscript{3} \citep{SrVO3-GW+DMFT-Jan-long}.
Inspired by this study, one can interpret the non-local static part
of the $GW$ self-energy as a screened-exchange term, while the dynamical
part is attributed to correlations:

\begin{equation}
\Sigma(k,\omega)=\Sigma_{sex}(k)+\Sigma_{c}(\omega)\label{eq:SEX and DMFT separation}
\end{equation}

This corresponds to the application of DMFT on a better starting point
than the usual DFT-LDA band structure, that is to say another quasiparticle
Hamiltonian $H_{0}$. In \citep{Ambroise-BaCo2As2,Ambroise-SrVO3},
$\Sigma_{sex}$ was approximated by a Hartree-Fock calculation in
which the Coulomb potential is replaced by a Yukawa potential defined
by the Thomas-Fermi screening length. $\Sigma_{c}$ is finally calculated
by Dynamical Mean Field Theory, including dynamical interactions $U(\omega)$.
This technique was dubbed Screened Exchange+Dynamical $U$ DMFT (SEx+DDMFT).

In general, non-local exchange results in an increase of the one-particle
bandwidth with respect to LDA, due to the reduced electronic repulsion.
On the other hand, the dynamical screening will reduce the bandwidth
by the bosonic renormalization factor $Z_{B}$ discussed above \citep{udyn-michele,udyneff-michele,udyn-werner}.
In BaCo$_{2}$As$_{2}$, the combination of those two effects results
in a decrease of the spectral weight at the Fermi level with respect
to the usual static-$U$ LDA+DMFT calculation. Moreover, the non-local
exchange modifies the shape of the Fermi surface, which is primordial
in such a compound with a flat band close to the Fermi level (see
Figure \ref{fig:BaCo2As2}) (and should also be in other pnictides
for the prediction of the low-energy properties). This makes the compound
fall just below the onset of the Stoner criterion. Similarly to what
happens with this weakly-correlated compound, we expect that a more
accurate treatment of non-local effects is also needed in other pnictides
(including the more correlated ones), in order to achieve a deeper
theoretical understanding (and, eventually, a first principles description)
of their physical properties.

\section{Conclusion}

We have reviewed some of the efforts of the last years that have been
concerned with the spectral properties of transition metal pnictide
compounds. During that period, PES and ARPES (and the preceding sample
elaborations that are a prerequisite to reaching good resolution)
have reached a mature stage where these techniques can be used to
study in some detail quasi-particle excitations and Fermi surfaces
in the normal phase, and provide useful information on gap structures
in the superconducting phase. At the same time, electronic structure
techniques for correlated materials have also tremendously evolved:
dynamical mean field theory-based calculations nowadays give quite
accurate descriptions of spectral, optical or magnetic properties.
Nevertheless, open questions also remain, namely at the interface
of electronic structure and many-body theory or concerning the limitations
of the local approximation done in DMFT. Due to their incredible sensitivity
to tiny changes in the Hund's exchange coupling, iron pnictides are
probably a kind of ``worst case scenario'' for electronic structure
calculations, but it is precisely for this reason that the field has
made so much progress over the last years: there are few materials
classes in nature where such a large number of relatively similar
compounds have been studied as intensively as in the field of transition
metal pnictides, providing a huge body of benchmark data and putting
serious constraints on theoretical modeling.

\section*{Acknowledgments}

Our view on the field has been shaped by discussions and collaborations
with numerous colleagues whom we thank warmly. This work was further
supported by supercomputing time at IDRIS/GENCI Orsay, under project
091393, an ERC Consolidator grant (project number 617196), NSF under
grant No. NSF PHY11- 25915, grants from MOST (2010CB923000, 2011CBA001000,
2011CBA00102, 2012CB821403 and 2013CB921703) and NSFC (11004232, 11034011/A0402,
11234014 and 11274362) from China, and by the Cai Yuanpei program.

\bibliographystyle{elsart-num}

\end{document}